\documentclass[lettersize,journal]{IEEEtran}
\usepackage{amsmath,amsfonts,amsthm,wasysym}
\usepackage{algorithmic}
\usepackage{algorithm}
\usepackage{array}
\usepackage{textcomp}
\usepackage{stfloats}
\usepackage{url}
\usepackage{verbatim}
\usepackage{graphicx}
\usepackage{cite}
\usepackage{subfigure}

\usepackage{bm}
\usepackage{comment}
\usepackage{xcolor}
\usepackage{soul}
\usepackage{url}

\usepackage{booktabs}

\usepackage{xspace}
\usepackage[inline]{enumitem}

\usepackage[colorinlistoftodos]{todonotes}

\usepackage{xspace}
\newcommand{\name}{DBS\xspace}

\newcommand{\revise}[1]{{\color{black}#1}}

\newcommand{\xp}{x^{\prime}{}}

\newcommand{\spr}{s^{\prime}{}}
\newcommand{\bc}{\bm{c}}
\newcommand{\bg}{\bm{g}}
\newcommand{\bk}{\bm{k}}
\newcommand{\br}{\bm{r}}
\newcommand{\brp}{\bm{r}^{\prime}{}}
\newcommand{\bv}{\bm{v}}
\newcommand{\bw}{\bm{w}}

\newcommand{\bx}{\bm{x}}

\newcommand{\by}{\bm{y}}
\newcommand{\bz}{\bm{z}}
\newcommand{\yp}{y^{\prime}{}}

\newcommand{\veceps}{\bm{\epsilon}}

\newcommand{\bara}{\overline{\alpha}}

\newcommand{\sct}{s_{\texttt{ct}}}

\newcommand{\wrap}{\mathcal{W}}
\newcommand{\ds}{\mathcal{D}}
\newcommand{\dsaug}{\mathcal{D}_{\texttt{aug}}}
\newcommand{\mg}{\widetilde{G}}
\newcommand{\loss}{\mathcal{L}}
\newcommand{\qcun}{q_{\theta}^{\text{ch }1}}
\newcommand{\qcdeux}{q_{\theta}^{\text{ch }2}}

\newcommand{\ppt}{\frac{\partial}{\partial \tau}}

\newcommand{\expn}[1]{\exp\left(#1\right)}
\newcommand{\norm}[1]{\left\lVert#1\right\rVert}
\newcommand{\wdiv}[2]{\text{W}_2\left(#1,#2\right)}

\newtheorem{lemma}{Lemma}[section]

\newtheorem{theorem}{Theorem}

\begin{document}

\title{Diffusion-Based Surrogate Modeling and Multi-Fidelity Calibration
}
\author{
    \IEEEauthorblockN{Naichen Shi\IEEEauthorrefmark{1}, Hao Yan\IEEEauthorrefmark{2}, Shenghan Guo\IEEEauthorrefmark{2}, Raed Al Kontar\IEEEauthorrefmark{3}}\\
    \IEEEauthorblockA{\IEEEauthorrefmark{1}Northwestern University}
    \IEEEauthorblockA{\IEEEauthorrefmark{2}Arizona State University}
    \IEEEauthorblockA{\IEEEauthorrefmark{3}University of Michigan
    alkontar@umich.edu}
}

\markboth{TASE}%
{Shell \MakeLowercase{\textit{et al.}}: A Sample Article Using IEEEtran.cls for IEEE Journals}


\maketitle

\begin{abstract}
\revise{Physics simulations have become fundamental tools to study myriad engineering systems. As physics simulations often involve simplifications, their outputs should be calibrated using real-world data. In this paper, we present a diffusion-based surrogate (\name) that calibrates multi-fidelity physics simulations with diffusion generative processes. \name categorizes multi-fidelity physics simulations into inexpensive and expensive simulations, depending on the computational costs. The inexpensive simulations, which can be obtained with low latency, directly inject contextual information into diffusion models. 
Furthermore, when results from expensive simulations are available, \name refines the quality of generated samples via a guided diffusion process. This design circumvents the need for large amounts of expensive physics simulations to train denoising diffusion models, thus lending flexibility to practitioners. 
\name builds on Bayesian probabilistic models and is equipped with a theoretical guarantee that provides upper bounds on the Wasserstein distance between the sample and underlying true distribution. The probabilistic nature of \name also provides a convenient approach for uncertainty quantification in prediction. Our models excel in cases where physics simulations are imperfect and sometimes inaccessible. We use a numerical simulation in fluid dynamics and a case study in laser-based metal powder deposition additive manufacturing to demonstrate how \name calibrates multi-fidelity physics simulations with observations to obtain surrogates with superior predictive performance.}
\end{abstract}

\def\abstractname{Note to Practitioners}
\begin{abstract}
\revise{In engineering applications, physics-based simulators are often employed to model complex systems. While these simulations encode our understanding of the underlying physics, they are frequently oversimplified or miscalibrated, leading to biased outputs. A natural approach to mitigating this bias is to calibrate simulation outputs using real-world data. Traditionally, Gaussian processes have been used for this purpose.
In this paper, we introduce an alternative calibration framework called Diffusion-based Surrogates (\name). \name leverages the flexibility of diffusion generative models to calibrate high-dimensional physics simulations. We introduce two designs to explicitly or implicitly incorporate physics simulations into the generative process. Our approach effectively integrates information from multi-fidelity physics models and excels in large-scale, high-dimensional calibration tasks. Notably, \name operates without requiring additional domain knowledge beyond simulation outputs. Further, \name is shown to effectively quantify the uncertainty in the predictions.}
\end{abstract}


\section{Introduction}
\IEEEPARstart{T}{he} 
era of generative AI is unfolding. Denoising diffusion process-based deep generative models, such as SORA~\cite{sora}, Midjourney~\cite{midjourney}, and Stable diffusion~\cite{stablediffusion}, can generate photorealistic and aesthetically pleasing images and videos with vivid details. At the heart of these generative models are score-based \textit{denoising diffusion models} (DDMs) designed to learn complex statistical patterns from high-dimensional training data~\cite{ddpm}. The flexible sampling procedure of DDMs allows for integrating prompts into the denoising process to generate controllable and customized samples~\cite{controlavideo}. 

Despite the success of DDMs in photo and video synthesis, two challenges hinder their application in engineering fields. First, the predictions generated by standard DDMs may not consistently align with the laws of physics. For example, even state-of-the-art DDMs like SORA can misinterpret physical interactions in the real world, generating videos that contain artifacts and lack long-term consistency. Second, DDMs are often data-hungry. To understand the detailed patterns in images, modern DDMs may need more than thousands of millions of training samples~\cite{clip}. This demand for large data limits the applicability of generative models in environments where acquiring high-quality and large-scale datasets is challenging or prohibitively expensive.

\revise{In the field of science and engineering, alternative solutions to pure data-driven statistical modeling exist.  Computer simulations based on physics principles naturally reflect underlying laws governing dynamic systems of interest. For example, in laser-based additive manufacturing, there has been remarkable progress in computer simulators that characterize the physics of meltpool, key hole, and thermal dynamics~\cite{gomelt,samaei2023multiphysics,tan2024multiphysics}. Compared with standard diffusion models, physical simulators often operate with a relatively small number of parameters and demonstrate interpretability and trustworthiness in both short and long-term predictions. Thus, a natural strategy is to embed such physical knowledge into generative models. Indeed, in natural language processing~\cite{nlpsurvey} and retrieval-augmented generation~\cite{rag}, a similar technique that combines factual knowledge with language generation, is prevalently applied~\cite{rag}.}

In the literature, two types of research areas are proposed to integrate physical knowledge into deep learning models. 1) \textbf{Physics-Informed Neural Networks (PINNs):
} PINNs learn the solution of a set of ordinary differential equations (ODEs) or partial differential equations (PDEs) by using neural networks (NNs), ensuring that the network's predictions are consistent with known physical principles. This approach typically involves incorporating differential equations that describe physical systems into the loss function of the NN~\cite{pinnreview}. Along this line of research, neural operators are designed to learn the nonlinear operators dictated by physical principles~\cite{deeponet,fourieroperator,neuraloperator}.
However, typical PINNs assume that the physical law is an accurate representation of the underlying system \cite{raissi2017physics,pinn}.  
2) \textbf{Physics-Constrained Neural Networks (PCNNs)}: PCNNs focus on enforcing physical or other knowledge constraints during the training process \cite{zhu2019physics, sun2020physics, zhang2020physics}. This can be achieved by adding regularization terms to the loss function, which penalizes deviations from known physical behaviors. Such constraints guide the learning process, ensuring that the resulting model adheres to physical principles. Along this line, there has been work on extending the constraints to Bayesian NNs to model and infer uncertainty in the data \cite{yang2020incorporating, huang2023posterior}. PCNNs are particularly useful when dealing with limited or noisy data, as the physical constraints help to regularize the learning process and prevent overfitting. However, the penalty approach also requires careful selection of the tuning parameters and assumes that the PDEs that characterize the evolution of the dynamic system are accurate to some degree. 

Despite the popularity of the models above, their outputs are vulnerable to model misspecifications because we rarely can easily model the exact underlying physics accurately or within a reasonable timeframe, especially for very complex processes commonly observed in engineering systems (see Section~\ref{sec:heat}). Instead, in many practical applications, physical knowledge is often implemented by computer simulators. 

\revise{In this paper, we directly analyze the outputs of such simulators. As simulators are frequently oversimplified or
miscalibrated, the outputs are often biased~\cite{simongraph}. Our overarching goal is to effectively \textbf{calibrate} these potentially biased outputs with real-world observations using DDMs to obtain improved surrogates capable of uncertainty quantification. A straightforward strategy is to take simulation outputs as additional inputs to the denoising neural network (DeNN) employed in the reverse diffusion process in DDMs. During the training stage, DeNNs are trained to gradually denoise corrupted samples under the guidance of physics simulations. The trained DeNNs then generate samples with the help of physics simulations in the inference stage.} 

Though intuitive and easy to implement, this approach has a critical caveat regarding computational costs. High-fidelity simulations require significant computing resources, such as modern numerical weather prediction programs, which use up to $10^{15}$ floating-point operations per second~\cite{nwpreview}. Simulators with lower computational demands can be available but often at the cost of worse performance~\cite{graphcast}. This trade-off between resources and performance is common across fields. 
Consequently, depending on the computational demands and resources available, results from expensive simulation programs may not be accessible in large scales.

Therefore, training data-hungry DeNNs on these results may not be feasible. We thus propose an alternative design to leverage the information from the potentially sparse simulation results. The method is inspired by conditional DDMs~\cite{songinverse}: instead of using expensive simulations as input, we use them to refine the sampling trajectory at the inference stage of the DDM. This strategy builds on Bayesian inference, which effectively borrows information from expensive simulations without retraining DeNNs.

\revise{Namely, we develop a diffusion-based surrogate model (\name) that calibrates multiple computer simulation models into with DDMs. In \name, we categorize physical simulators into two classes: inexpensive simulators, whose results are easily obtainable with low latency, and expensive simulators, which output results with higher fidelity but consume larger computational resources. We design separate knowledge integration techniques for different simulators. More specifically, we use inexpensive computer simulations as additional inputs to the DeNN, which is trained to generate predictions with insights from simulations. For the expensive computer simulations, we construct a separate conditional probability model and use a conditional diffusion process to further improve the sampling quality. The design of \name decouples the training of probabilistic models for different physics simulators, facilitating its implementation in practice, especially when some simulation results are not always available.}

The proposed \name is a hybrid of physics-based computer simulations and data-driven DDMs, thus reaping advantages from both worlds. The model is a physics-informed surrogate~\cite{surrogates} that inherits physics knowledge from simulations while learning statistical patterns from data. As a result, the generated predictions can abide by the principles of physics while remaining consistent with the observations.

We highlight several benefits of \name. \textit{Generality}: \name operates on the outputs of simulators rather than specific forms of ODEs or PDEs. Thus, \name can work with a wide range of physics simulators, even if the simulators are black-box functions for practitioners. \textit{Flexibility}: The conditional probability models for different physics simulators can be trained separately. Such a decoupled design provides an interface that allows for easy plug-ins of the results from different simulators. \textit{Computational efficiency}: Unlike physics surrogates implemented by Gaussian processes~\cite{surrogates}, whose computation complexity often scales quadratically or even cubically with the training dataset size~\cite{gp4ml}, the inference time complexity of \name is independent of the training dataset size. Additionally, the DNN in \name shows strong performance in modeling high-dimensional data in practice.

We demonstrate the capability of \name on two exemplary applications: a \textit{fluid system} and a \textit{thermal process} from additive manufacturing, each representing different types of physics simulations. 
Both applications illustrate how \name integrates statistical knowledge from real observations with physics knowledge from simulations to better predict the evolution of physical systems. The code to reproduce numerical results in this paper is available in the repository {\color{blue}\url{https://github.com/UMDataScienceLab/MGDM}}.

\revise{We summarize our contributions in the following,
\begin{itemize}
    \item From a simulation and calibration perspective, we propose \name that introduces sampling-based approaches to calibrate high-dimensional physics simulation outputs by diffusion models based on conditional denoising neural networks and conditional inverse diffusion processes. The proposed approaches are flexible, scalable to high dimensions, and equipped with uncertainty quantification capabilities. We also provide a theoretical guarantee for the sampling distributions.
    \item On the computation side, we introduce two designs of energy-based guidance from simulation outputs in the conditional reverse diffusion process. We also propose an efficient approximation to the conditional score function, significantly reducing memory consumption in gradient estimation.
    \item From an application perspective, we showcase the effectiveness of \name in laser-based metal additive manufacturing (LBMAM) process characterization. We develop two ad-hoc physics simulators for meltpool thermal dynamics and spatter movement. Results show that both simulators integrate seamlessly with our proposed \name framework. Moreover, the sampling quality improves as we incorporate more physics information in \name. 
\end{itemize}
}

\section{Related work}
This section delves into recent advancements and applications of denoising diffusion models (DDM)s, particularly in the context of video generation and integration of physics simulations. We introduce the basics of DDMs, their conditional and constrained counterparts, the use of physics surrogates to enhance performance, and the embedding of physics knowledge into the diffusion process to highlight our framework's methodology and benefits.


\paragraph{Diffusion and video generation} \label{subsec: diffvideo} DDMs ~\cite{ddpm} introduce a flexible and expressive framework for generative models. The connections between DDMs, score matching, and stochastic differential equations are explored in a series of works~\cite{ddim,song2019generative,songkldiv}. As discussed, DDMs form the backbone of multiple modern large-scale generative models~\cite{sora,midjourney,stablediffusion}.

\revise{Significant recent efforts have been made to improve the performance and efficiency of DDMs. The current state-of-the-art Frechet Inception Distance (FID)~\cite{fid} on ImageNet is achieved by~\cite{imagenetsota}, featuring techniques including latent diffusion models~\cite{ldm}, which integrate dimensionality reduction with diffusion processes, and DiT~\cite{dit}, a vision transformer-based architecture designed for large-scale diffusion model training. DiT serves as the backbone for several foundational diffusion models, including WALT~\cite{walt}, SORA~\cite{sora}, Stable Diffusion~\cite{stablediffusion}, and DALL$\cdot$E~\cite{dalle}. Block diffusion~\cite{blockdiffusion} and LLaDA
~\cite{llmdm} combines DM with the auto-regressive modeling in LLMs. Beyond neural network architecture advancements, recent works have also explored novel training and inference strategies to further improve generative performance. ~\cite{flowmatching} uses a flow-matching objective that learns the optimal sampling path. ~\cite{repalign} proposes a representation alignment regularization to improve the training process. Also, DPM~\cite{dpm} aims to accelerate the diffusion sampling process through novel discretization schemes. Many improvements can be readily integrated into the \name framework.}

In the temporal data generation domain, DDM and its variants can effectively model temporal interactions in multivariate time series~\cite{timeseries1,timeseries2} and video frames~\cite{imagen,videodiffusion}. Our work also predicts temporal evolutions of dynamic systems, but under the guidance of physics simulations.

\paragraph{Conditional diffusion} \label{subsec: conddiff}
 Recent methods have been proposed to leverage the information from external conditions to guide diffusion processes~\cite{saharia2022image,chung2023guide,diffusionsurvey,accelerate}. A well-known example of this practice is the use of spectral signals in the frequency domain to inform Magnetic Resonance Imaging (MRI) reconstruction~\cite{jalalinverse,songinverse}. In video generation, motion vectors can also guide the spatial and temporal evolution of frames~\cite{videocomposer}. Modern video generative models often condition on input texts as well~\cite{controlavideo,sora}. With a similar rationale, we condition on physics simulators to guide DDMs. 

\paragraph{Physics-informed surrogates} \label{subsec: physics}
Research that aims to combine statistical models with physics knowledge has a long history. One prominent method, proposed by Kennedy and O'Hagan (KOH)~\cite{kennedy2001bayesian,kennedy2000predicting}, predicts the discrepancy between physics simulations and real-life observations by using GPs and employs a Bayesian calibration approach to optimize the parameters. In the literature of experimental design, such statistical models that emulate physics observations or computer simulations are often called surrogate models~\cite{surrogates}. \revise{Many multi-fidelity surrogate models build on GPs~\cite{jeffsurrogate,simongraph,wusinglemeshindex,gpcv,physicsprior}. Among them, \cite{wusinglemeshindex} uses the fidelity parameter as a contextual input to the GP, and \cite{ji2024conglomerate} extends the framework to the multivariate fidelity parameter setting. \cite{simongraph} builds a graph that connects lower fidelity models with high fidelity ones. \cite{nonhierarchicalfusion} considers the setting where the fidelity index is unknown. \cite{gpcv} calibrates the GP variance prediction by leave-one-out cross validation.} \revise{Different from existing approaches, \name does not make structural assumptions on the correlations between simulations and experimental observations. Instead, we train a neural network to automatically capture such correlations. Also, the application of GPs in large-scale and high-dimensional datasets is limited~\cite{gp4ml}. Our proposed model \name circumvents the issue by using DDMs.}


\paragraph{Physics-driven diffusion} \label{subsec: physicsdiff}
A few recent works propose to bring physics knowledge into DDMs~\cite{conditionaldiffusionfluid,cocogen,molinaro2024generative}. Among them, CoCoGen~\cite{cocogen} enforces PDE constraints onto the reverse diffusion process, which improves the performance of Darcy flow modeling. However, in broader applications, imposing PDE constraints can be too restrictive and exacerbate modeling bias~\cite{surrogates}.  \revise{GenCFD~\cite{molinaro2024generative} establishes the advantage of DDMs theoretically but does not leverage the information from simulations at the inference stage.} \cite{conditionaldiffusionfluid} uses residuals of the PDE as additional inputs to the denoising network and achieves remarkable performance in fluid field super-resolution. Despite the success, it is uneasy to apply the method in~\cite{conditionaldiffusionfluid} to applications where the physics cannot be described by a single PDE. Unlike these approaches, our method \name does not require the knowledge of the underlying PDE. As long as practitioners have access to the outputs of physics simulators, they can use the outputs to guide their DDMs. Hence, the requirement for domain knowledge is minimized. Additionally, since the \name is trained on real observations, the statistical knowledge from data can be leveraged to improve the results from potentially biased physics simulations. 
\revise{Concurrent to our work, \cite{liu2024physgen} also refines simulation-generated videos. However, it does not handle multi-fidelity simulation.}

\paragraph{Constrained diffusion} \label{subsec: constraindiff} Constrained DDMs have been extensively studied to understand how physical constraints influence the training process of DDMs. These studies reveal that incorporating physical constraints, such as boundaries or barriers, significantly alters diffusion dynamics compared to unrestricted environments. Recent advancements include DDMs on Riemannian manifolds~\cite{de2022riemannian,huang2022riemannian}, investigating the diffusion dynamics on Riemannian manifolds,  and DDMs on constrained domains~\cite{fishman2023diffusion}, introducing the logarithmic barrier metric and reflected Brownian motion, demonstrating practical utility in fields like robotics and protein design. Constrained DDMs have found applications in robotics~\cite{se3} and crystal structure prediction~\cite{diffusioncrystal}. However, these methods often assume that the constrained domain for the DDMs is known, which is challenging to determine in complex real-world systems.

\section{Model}
In this section, we progressively construct models that fuse knowledge from multiple physics simulators into DDMs. The overreaching goal is to predict the evolution of dynamical systems to high fidelity and verisimilitude with historical observations and access to physics simulations. We use a vector $\bx_{0,s}$ to represent the state of the system at time $s$, where the subscript $0$ denotes the real-life observed data. 

\begin{figure}[h]
    \centering
    \includegraphics[width=0.45\textwidth]{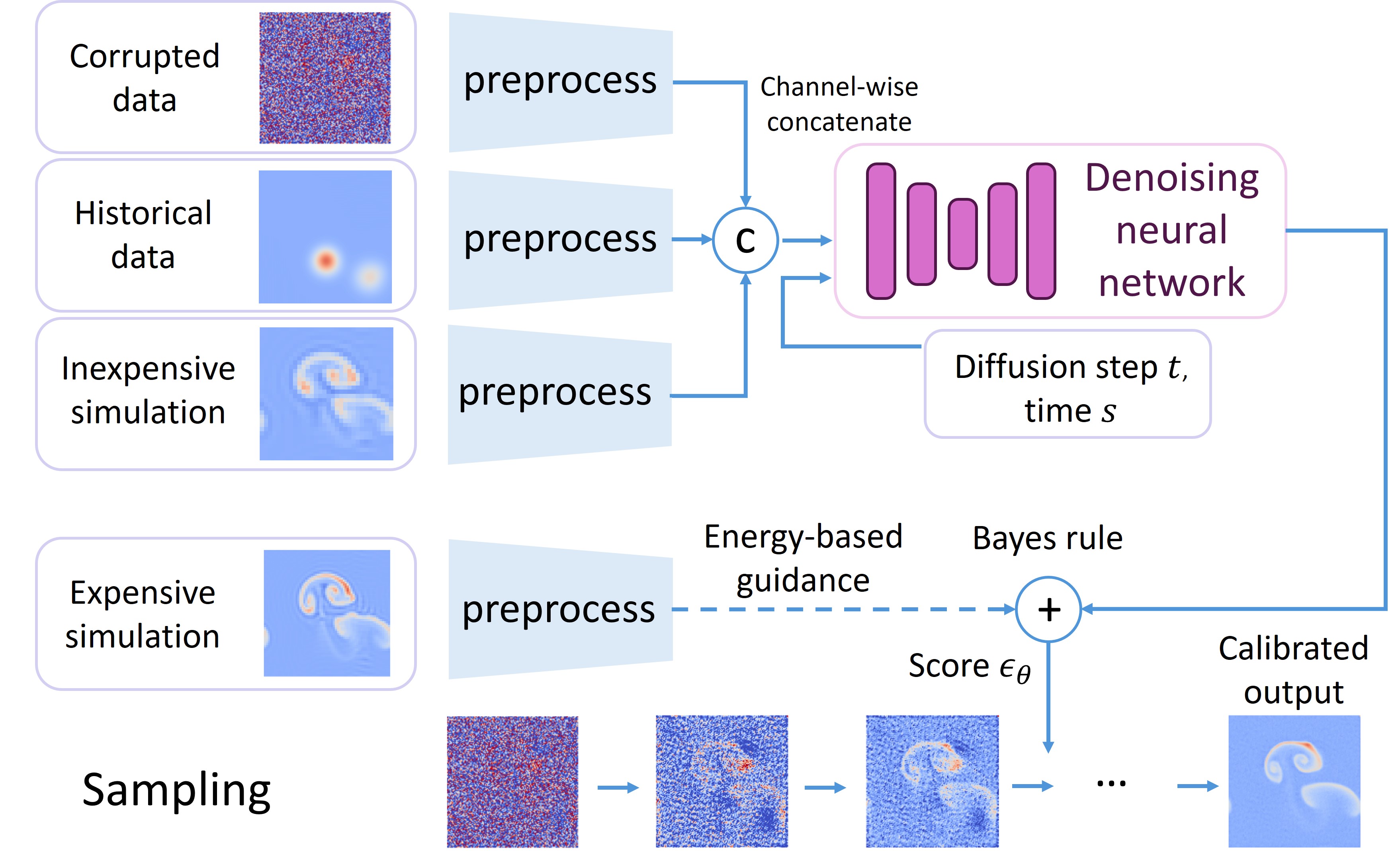}
    
    \caption{\revise{A schematic plot of the \name framework. Dashed arrow means optional conditioning from expensive simulation.}}
    \label{fig:scheme_plot}
\end{figure}

\revise{As discussed, we group multi-fidelity physical simulators into two categories: the inexpensive simulation and the expensive simulation. We assume the practitioner can call inexpensive simulators at each time step $s$ at low latency. The result is denoted as a vector  $\bc_{s,1}$. Expensive physics simulations are more time-consuming and may not be accessible at all $s$. We use a vector $\bc_{s,2}$ to denote the result of expensive simulation at time $s$ if available. The generic notions of $\bc_{s,1}$ and $\bc_{s,2}$ can incorporate a broad range of computer simulations or statistical surrogates~\cite{surrogates}. The framework of \name is plotted in Fig.~\ref{fig:scheme_plot}. }

In the rest of this section, we will first review the DDM framework with a focus on denoising diffusion implicit models (DDIMs)~\cite{ddim}, an instance of DDMs popular in the field of image and video generation~\cite{ldm,diffusionsurvey}.  
We will then explain the details in \name and develop techniques to incorporate outputs from both inexpensive and expensive computer simulations. Finally, we will present the pseudocode for our training and sampling algorithms.

\subsection{Standard diffusion model}
\label{sec:standarddiff}
Denoising Diffusion Implicit Models (DDIMs) use a series of Gaussian noise with increasing variance to corrupt the data, then train Denoising Neural Networks (DeNNs) to gradually reconstruct clean data from corrupted ones. The DDIM consists of multiple steps $t=0,1,2,\cdots,T$~\cite{ddpm}, each one of which corresponds to a specific level of variance in the Gaussian noise. For clarity, we use $\bx_{t,s}$ to denote the step-$t$ diffusion of the state vector observed at time $s$. It is important to note that $t$ signifies the diffusion step, whereas $s$ indicates the actual time within the dynamic system.

In DDIM, the forward diffusion process is given as,
\begin{equation}
\label{eqn:forwarddiffusion}
\bx_{t,s} = \sqrt{1-\beta_t} \, \bx_{t-1,s} + \sqrt{\beta_t} \, \bz_{t,s},
\end{equation}
where $\bz_{t,s}$ are i.i.d. Gaussian noise vectors and $\beta_t$ is a predefined constant that determines the noise variances. Similar to~\cite{ddpm}, we introduce notation $\alpha_t=1-\beta_t$ and $\bara_t=\prod_{\tau=1}^t \alpha_{\tau}$.

In the continuous limit, the forward diffusion process~\eqref{eqn:forwarddiffusion} reduces to a stochastic differential equation (SDE)~\cite{song2019generative},
\begin{equation}
\label{eqn:forwardsde}
d\bx_{t,s} = -\frac{\beta_t}{2}\bx_{t,s}dt + \sqrt{\beta_t}d\bw_t,
\end{equation}
where $\bw_t$ is the standard Brownian motion or Weiner process. With a slight abuse of notation, the subscript $t$ denotes a continuous variable in~\eqref{eqn:forwardsde} and a discrete variable in~\eqref{eqn:forwarddiffusion}. In literature,~\eqref{eqn:forwardsde} is often called a variance-preserving SDE~\cite{song2019generative}.

In the task of future physics state prediction, historic observations are often available to practitioners. We use $\sct$ to denote the number of context (already observed) observations, and $\bx_{0,1:\sct}$ as a shorthand notation for the concatenated observed vector $[\bx_{0,1},\bx_{0,2},\cdots,\bx_{0,\sct}]$. Mathematically, the task can be formulated as predicting/sampling the state vector $\bx_{0,s^{*}}$ at a time of interest $s^*$ from the predictive distribution $p(\bx_{0,s^{*}}|\bx_{0,1:\sct})$, where $\sct<s^{*}$. The target distribution $p(\bx_{0,s^{*}}|\bx_{0,1:\sct})$ is a conditional distribution of future state vector $\bx_{0,s^{*}}$ given the observed state vectors $\bx_{0,1:\sct}$. 

It is important to note that for predictions at multiple future time points, or multiple values of $s^{*}$, practitioners can efficiently apply the DDM framework by simply stacking these state vectors and sampling from the the distribution of stacked vectors. Hence, for simplicity and without any loss of generality, we use $\bx_{0,s^{*}}$ to signify the state vector at any given target time $s^{*}$. 

To generate high-quality samples, DDMs exploit an existing dataset $\ds = \{(\bx_{0,1:\sct}^{(i)},\bx_{0,s^{*}}^{(i)})\}_{i=1}^N$ to learn the target conditional distribution, where the superscript $(i)$ is the observation index. Different $i$ denotes different collected evolution trajectories of state vectors. $N$ is the total number of trajectories in the training set.

We briefly describe the training objective of DDIM. By iteratively applying~\eqref{eqn:forwarddiffusion}, one can show that $\bx_{t,s}$ has the same distribution as $\sqrt{\overline{\alpha}_t}\bx_{0,s^{*}}+\sqrt{1-\overline{\alpha}_t}\veceps$ where $\veceps$ is a vector whose elements are i.i.d. standard Gaussians. DDIM leverages such fact to train an iterative DeNN $\veceps_{\theta}(\cdot)$ that predicts the noise $\veceps$ from the corrupted sample $\bx_{t,s^{*}}$. More specifically, the training objective of DDIM is,
\begin{align}
\label{eqn:standarddiffisiontrain}
\min_{\theta} &\quad \mathbb{E}_{(\bx_{0,s^{*}},\bx_{0,1:\sct})\sim \ds,\,t\sim \mathcal{U}[0,T],\,\veceps\sim \mathcal{N}(0,\mathbf{I})}\notag\\
&\left[\norm{\veceps-\veceps_{\theta}(\sqrt{\overline{\alpha}_t}\bx_{0,s^{*}}+\sqrt{1-\overline{\alpha}_t}\veceps,t,\bx_{0,1:\sct})}^2\right].
\end{align}
In~\eqref{eqn:standarddiffisiontrain}, the diffusion and denoising only happen at the target time $s^{*}$. The objective is to minimize the difference between the noise added to the sample and the noise predicted by the denoising network. 
It is worth noting that theoretically, the denoising objective~\eqref{eqn:standarddiffisiontrain} is related to the score function in statistics: roughly speaking, if the sample size goes to infinity and~\eqref{eqn:standarddiffisiontrain} is exactly minimized, the optimal $\veceps_{\theta}^{\star}$ becomes~\cite{song2019generative},
\begin{equation}
\label{eqn:epsilonsolution}
\veceps_{\theta}^{\star}(\bx_{t,s},t,\bx_{0,1:\sct}) = -\sqrt{1-\bara_t}\nabla_{\bx_{t,s}} \log p\left(\bx_{t,s^{*}}|\bx_{0,1:\sct}\right),
\end{equation}
where $p\left(\bx_{t,s^{*}}|\bx_{0,1:\sct}\right)$ is the p.d.f. of the random vector $\bx_{t,s^{*}}=\sqrt{\overline{\alpha}_t}\bx_{0,s^{*}}+\sqrt{1-\overline{\alpha}_t}\veceps$, and the score function is the gradient of the logarithm of the p.d.f.

In the inference stage, DDIM~\cite{ddim} generates high-quality samples from the approximated score functions. More precisely, DDIM samples $\bx_{T,s^{*}}$ from the standard normal distribution, then applies the denoising network $\veceps_{\theta}$ to iteratively denoise $\bx_{t,s^{*}}$: 
\begin{align}
\label{eqn:standarddiffusionsample}
&\bx_{t-1,s^{*}} = \frac{1}{\sqrt{\alpha_t}}\Big(\bx_{t,s^{*}}-\frac{\left(1-\alpha_t\right)\veceps_{\theta}(\bx_{t,s^{*}},t,\bx_{0,1:\sct})}{\sqrt{1-\overline{\alpha}_t}+\sqrt{\alpha_t-\bara_t}}\Big),
\end{align}
for $t$ from $T$ to $1$. The coefficient $\frac{1-\alpha_t}{\sqrt{1-\overline{\alpha}_t}+\sqrt{\alpha_t-\bara_t}}$ will converge to $\frac{\beta_t}{2\sqrt{1-\bara_t}}$ when $\beta_t$ is small. We obtain $\bx_{0,s^{*}}$ eventually. 

From the perspective of SDEs, if the denoising network is properly trained as in~\eqref{eqn:epsilonsolution}, the sampling rule~\eqref{eqn:standarddiffusionsample} is a discretized version of the reverse process ODE~\eqref{eqn:forwardsde},
\begin{equation}
\label{eqn:reversesde}
d\bx_{t,s^{*}} = \left(\frac{\beta_t\bx_{t,s^{*}}}{2}+\frac{\beta_t}{2}\nabla_{\bx_{t,s^{*}}} \log p(\bx_{t,s^{*}}|\bx_{0,1:\sct})\right)dt. 
\end{equation}
Under the dynamics specified by~\eqref{eqn:reversesde}, the random vector $\bx_{0,s^{*}}$ will follow the desired predictive distribution $p(\bx_{0,s^{*}}|\bx_{0,1:\sct})$ if properly initialized. Such connection justifies~\eqref{eqn:standarddiffusionsample} theoretically~\cite{ddim}.

\subsection{Inexpensive physics-conditioned diffusion}
\label{sec:physicsguidedmodel}
As discussed, the training objective~\eqref{eqn:standarddiffisiontrain} and sampling scheme~\eqref{eqn:standarddiffusionsample} (or the continuous version~\eqref{eqn:reversesde}) only focus on the statistical patterns in the data and can overlook the physics mechanisms, especially when the size of the training dataset $\ds$ is not extremely large. Physics simulations can help alleviate the issue. We assume that simulations can make predictions about the future evolution of the system. The output at time $s$ is $\bc_{1,s}$. We further assume here that the physics simulations are inexpensive, allowing simulation predictions to be obtained for each sample in the training and sampling stages with low latency.

With the physics simulator, we can build an augmented training dataset by combining the training data and inexpensive simulation data at target time $s^{*}$, $\dsaug=\{( \bx_{0,s^{*}}^{(i)},\bx_{0,1:\sct}^{(i)},\bc_{1,s^{*}}^{(i)})\}_{i=1}^N$. The augmented dataset enables us to train a conditional diffusion network $\veceps_{\theta}(\cdot)$ that takes not only the initial frames $\bx_{0,1:\sct}$ but also the simulation prediction $\bc_{1,s^{*}}$ as its contextual input.

The training objective of the inexpensive physics-conditioned diffusion model thus becomes
\begin{align}
\label{eqn:inexpensivediffusiontrain}
&\min_{\theta} \mathbb{E}_{(\bx_{0,s^{*}},\bx_{0,1:\sct},\bc_{1,s^{*}})\sim \dsaug,t\sim \mathcal{U}[0,T],\veceps\sim\mathcal{N}(0,\mathbf{I})}\Big[\notag\\
&\norm{\veceps-\veceps_{\theta}(\sqrt{\overline{\alpha}_t}\bx_{0,s^{*}}+\sqrt{1-\overline{\alpha}_t}\veceps,t,\bx_{0,1:\sct}, \bc_{1,s^{*}})}^2\Big].
\end{align}
Similar to DDIM, we use Monte Carlo approximation to minimize the objective. The pseudocode is presented in Algorithm~\ref{alg:trainepsilon}.

Intuitively, the physics context $\bc_{1,s^{*}}$ can bring additional physics knowledge to the model, thus augmenting the authenticity of the prediction. The denoising network $\veceps_{\theta}$ is then trained to absorb such physics knowledge. Such intuition is corroborated by our theoretical analysis in Theorem~\ref{thm:wassdis} (see Section~\ref{sec:theory}).

Accordingly, the iterative sampling rule becomes,
\begin{align}
\label{eqn:inexpensivediffusionsample}
&\bx_{t-1,s^{*}} = \frac{1}{\sqrt{\alpha_t}}\Big(\bx_{t,s^{*}}-\frac{\left(1-\alpha_t\right)\veceps_{\theta}(\bx_{t,s^{*}},t,\bx_{0,1:\sct},\bc_{1,s^{*}})}{\sqrt{1-\overline{\alpha}_t}+\sqrt{\alpha_t-\bara_t}}\Big),  
\end{align}
for $t$ from $T$ to $1$, starting from $\bx_{T,s^{*}}\sim\mathcal{N}(0,\mathbf{I})$.  

The sampling rule~\eqref{eqn:inexpensivediffusionsample} is analogous to that in~\eqref{eqn:standarddiffusionsample}. 
The only difference is that we augment the input to the DeNN with the predictions from the physics simulations. We also provide a pseudocode of the sampling rule as the choice $1$ of Algorithm~\ref{alg:sample}. Our experiments show that the augmented information can significantly improve the sampling performance. 


\subsection{Expensive physics-conditioned diffusion}
\label{sec:doublephysicsguidedmodel}
Clearly, the approach above is simple since simulations are cheap, but what if we have simulations that are expensive? Often, practitioners can obtain physics predictions from more expensive but potentially more accurate physics models. We denote the results from an expensive simulator at time $s^{*}$ as $\bc_{2,s^{*}}$. Then, an augmented dataset of trajectories and simulations is $\{\bx_{0,s^{*}}^{(i)},\bx_{0,1:\sct}^{(i)},\bc_{1,s^{*}}^{(i)},\bc_{2,s^{*}}^{(i)}\}_{i=1}^N$. However, due to high computation costs or latency, we assume that ${\bc_{2,s^{*}}^{(i)}}$ may only be available for a subset of $i\in\mathcal{S}_{\text{available}}$. Thus, we cannot directly feed $\bc_{2,s^{*}}$ into the DeNN. This section employs a different strategy to handle the case where $\bc_{2,s^{*}}^{(i)}$ is available. Here, we leverage conditional DDMs to guide the diffusion process by both inexpensive and expensive simulators, namely, $p(\bx_{0,s^{*}}|\bc_{1,s^{*}}, \bc_{2,s^{*}})$. Notably, our method exploits the sparsely available $\bc_{2,s^{*}}$, decoupled from the training procedure in Section~\ref{sec:physicsguidedmodel}, ensuring that the unavailability of $\bc_{2,s^{*}}^{(i)}$ does not affect the training of the denoising network $\veceps_{\theta}$.

We motivate our derivation from the reverse diffusion ODE,
\begin{align}
\label{eqn:conditionalreversesde}
&d\bx_{t,s^{*}}\notag \\
&= \frac{\beta_t}{2}\left(\bx_{t,s^{*}}+\nabla_{\bx_{t,s^{*}}} \log p(\bx_{t,s^{*}}|\bx_{0,1:\sct},\bc_{1,s^{*}},\bc_{2,s^{*}})\right)dt.
\end{align}

One can see that the conditional score function $\nabla_{\bx_{t,s^{*}}}\log p(\bx_{t,s^{*}}|\bx_{0,1:\sct},\bc_{1,s^{*}},\bc_{2,s^{*}})$ replaces its counterpart in~\eqref{eqn:reversesde} and plays the central role in the reverse diffusion process. Therefore, it suffices to derive an estimate of the conditional score function.

From the Bayes's rule, we know for any $t\ge 0$,
\begin{align*}
&\nabla \log p(\bx_{t,s^{*}}|\bc_{1,s^{*}},\bc_{2,s^{*}})\\
&= \nabla\log p(\bx_{t,s^{*}}|\bc_{1,s^{*}}) + \nabla\log p(\bc_{2,s^{*}}|\bx_{t,s^{*}},\bc_{1,s^{*}}).  
\end{align*}

The first term can be approximated by the denoising network trained by Algorithm~\ref{alg:trainepsilon}.  We use $\bg(\bx_{0,1:\sct}, \bc_{2,s^{*}},\bc_{1,s^{*}},t)$ to denote an estimate of the second term,
\begin{align}
\label{eqn:gdef}
 &\bg(\bx_{0,1:\sct}, \bc_{1,s^{*}},\bc_{2,s^{*}},t)\approx\nabla_{\bx_{t,s^{*}}} \log  p(\bc_{2,s^{*}}|\bx_{t,s^{*}},\bc_{1,s^{*}}) .  
\end{align}
In literature, there are multiple ways to construct estimates for $\bg$. When the conditional probability $p(\bc_{2,s^{*}}|\bx_{0,s^{*}},\bc_{1,s^{*}})$ is known, we introduce a conceptually simple and computationally tractable procedure inspired by~\cite{chung2023guide}. The pseudocode is presented in Algorithm~\ref{alg:gdef}.

\begin{algorithm}
\caption{Estimate the gradient of the log conditional probability $\nabla_{\bx_{t,s^{*}}} \log  p(\bc_{2,s^{*}}|\bx_{t,s^{*}},\bc_{1,s^{*}})$}
\label{alg:gdef}
\begin{algorithmic}[1]
\STATE Input the conditional distribution $p(\bc_{2,s^{*}}|\bx_{0,s^{*}},\bc_{1,s^{*}})$,
\STATE \label{step:tweedie}Estimate $\hat{\bx}_{0,s^{*},\theta}(\bx_{t,s^{*}},\bc_{1,s^{*}},t)=  \frac{\bx_{t,s^{*}}-\sqrt{1-\bara_t}\veceps_{\theta}(\bx_{t,s^{*}},\bc_{1,s^{*}},t)}{\sqrt{\bara_t}}$,
    \STATE \label{step:nablax0}Calculate $\nabla_{\hat{\bx}_{0,s^{*},\theta}}\log p(\bc_{2,s^{*}}|\hat{\bx}_{0,s^{*},\theta},\bc_{1,s^{*}})$,
    \STATE \label{step:nablaxt}Take $\bg=\frac{1}{\sqrt{\bara_t}}\nabla_{\hat{\bx}_{0,s^{*},\theta}}\log p(\bc_{2,s^{*}}|\hat{\bx}_{0,s^{*},\theta},\bc_{1,s^{*}})$.

\STATE Return $\bg$.
\end{algorithmic}
\end{algorithm}

In Algorithm~\ref{alg:gdef}, step~\ref{step:tweedie} uses  the Tweedie’s formula~\cite{chung2023guide} to produce a point estimate of the sample $\hat{\bx}_{0,s^{*},\theta}$ given a noisy sample $\bx_{t,s^{*}}$. It essentially removes the noise from the noisy sample with the noise estimated by the DeNN $\veceps_{\theta}$. Then, step~\ref{step:nablax0} and~\ref{step:nablaxt} use the (scaled) gradient over the clean sample $\hat{\bx}_{0,s^{*},\theta}$ to approximate $\nabla_{\bx_{t,s^{*}}} \log  p(\bc_{2,s^{*}}|\bx_{t,s^{*}},\bc_{1,s^{*}})$. Such a procedure is easy to implement and works well in practice. We will defer detailed derivations to supplementary materials.

It is the practitioner's discretion to choose the instantiation of $p(\bc_{2,s^{*}}|\bx_{0,s^{*}},\bc_{1,s^{*}})$ in Algorithm~\ref{alg:gdef}. In principle, the conditional probability model should reflect how the expensive simulation $\bc_{2,s^{*}}$ is related to the observations $\bx_{0,s^{*}}$. We will demonstrate two choices of conditional probability models in the numerical experiments about fumes~\eqref{eqn:patchmodel} and thermal processes~\eqref{eqn:flowphysicsprobability}. In section~\ref{sec:guideline}, we also present some guidelines for designing the conditional model.


Combining Algorithm~\ref{alg:gdef} with the DDIM discretization of~\eqref{eqn:conditionalreversesde}, a discrete update rule for sampling/inference is,
\begin{align}
&\bx_{t-1,s^{*}} = \frac{1}{\sqrt{\alpha_t}}\Big(\bx_{t,s^{*}}\tag{Term 1}\label{eqn:term1}\\
&-\frac{1-\alpha_t}{\sqrt{1-\overline{\alpha}_t}+\sqrt{\alpha_t-\bara_t}}\veceps_{\theta}(\bx_{t,s^{*}},t,\bx_{0,1:\sct}, \bc_{1,s^{*}})\tag{Term 2}\label{eqn:term2}\\
&+(1-\alpha_t)\bg(\bx_{0,1:\sct}, \bc_{1,s^{*}},\bc_{2,s^{*}},t)\Big),\tag{Term 3}\label{eqn:term3}\\
\label{eqn:expensivediffusionsample}
\end{align}
for $t$ from $T$ to $1$. In~\eqref{eqn:expensivediffusionsample}, ~\eqref{eqn:term1} corresponds to the $\frac{\beta_t}{2}\bx_{t,s^{*}}$ component in~\eqref{eqn:conditionalreversesde}, which is essential in the variance-preserving SDE. ~\eqref{eqn:term2} approximates the score function that drives the sample $\bx_{t,s^{*}}$ to high-probability regions predicted by the inexpensive physics simulation $\bc_{1,s^{*}}$. The coefficients are consistent with those in DDIM. Furthermore, ~\eqref{eqn:term3} represents the conditioning of the expensive physics simulation $\bc_{2,s^{*}}$ that furnishes additional guidance to $\bx_{t,s^{*}}$. ~\eqref{eqn:term3} is the major difference between~\eqref{eqn:expensivediffusionsample} and~\eqref{eqn:inexpensivediffusionsample}. The collective effects of three forces encourage the sample to enter regions where statistical patterns and physics knowledge are congruent. 

By initializing $\bx_{T,s^{*}}$ from multiple independent samples from the standard normal distribution and iteratively applying~\eqref{eqn:expensivediffusionsample}, one can obtain multiple instances of $\bx_{0,s^{*}}$. The sample variances estimated from these instances provide a straightforward characterization for uncertainty quantification.

To summarize, the pseudocodes of the training and sampling algorithms are presented in Algorithm~\ref{alg:trainepsilon} and Algorithm~\ref{alg:sample}.
\begin{algorithm}
\caption{\name: Training of the denoising network $\veceps_{\theta}$}
\label{alg:trainepsilon}
\begin{algorithmic}[1]
\STATE Input training dataset $\dsaug$.

\FOR{
Epoch $n=1,2,\cdots,B$}
\STATE Sample $\veceps\sim\mathcal{N}(0,\mathbf{I})$.
\STATE Sample $\left(\bx_{0,s^{*}},\bx_{0,1:\sct},\bc_{1,s^{*}}\right)\sim\dsaug$.
\STATE Sample $t\sim\mathcal{U}[0,T]$.
\STATE Calculate the gradient\\ $\nabla_{\theta}\norm{\veceps-\veceps_{\theta}(\sqrt{\bara_t}\bx_{0,s^{*}}+\sqrt{1-\bara_t}\veceps,\bx_{0,1:\sct},\bc_{1,s^{*}},t)}^2$.
\STATE Update $\theta$ by the gradient.
\ENDFOR
\STATE Return $\veceps_{\theta}$.
\end{algorithmic}
\end{algorithm}

\begin{algorithm}
\caption{\name: Sample from $p(\bx_{0,s^{*}}|\bx_{0,1:\sct},\bc_{1,s^{*}},\bc_{2,s^{*}}$.)}
\label{alg:sample}
\begin{algorithmic}[1]
\STATE Input trained denoising network $\veceps_{\theta}$, $\bg$, context state vectors $\bx_{0,1:\sct}$, inexpensive physics output $\bc_{1,s^{*}}$, (perhaps) expensive physics output $\bc_{2,s^{*}}$.
\STATE Sample $\bx_{T,s^{*}}\sim\mathcal{N}(0,\mathbf{I})$.
\FOR{
Index $t=T,T-1,\cdots,1$}
\IF{(Choice 1) $\bc_{2,s^{*}}$ is not available}
\STATE Calculate $\bx_{t-1,s^{*}}$ from~\eqref{eqn:inexpensivediffusionsample}.
\ENDIF
\IF{(Choice 2) $\bc_{2,s^{*}}$ is available }
\STATE Calculate $\bx_{t-1,s^{*}}$ from~\eqref{eqn:expensivediffusionsample}.
\ENDIF
\ENDFOR
\STATE Return $\bx_{0,s^{*}}$.
\end{algorithmic}
\end{algorithm}

\subsection{Insights for designing the conditional probability $p(\bc_{2,s^{*}}|\bx_{0,s^{*}},\bc_{1,s^{*}})$ }
\label{sec:guideline}
Leveraging domain-specific knowledge is crucial for determining the exact form of the conditional probability model. In literature,  numerous successful examples of such models exist~\cite{cocogen, songinverse}.

In scenarios where different physics simulations are independent, it is reasonable to simplify the conditional probability as \( p(\bc_{2,s^{*}}|\bx_{0,s^{*}},\bc_{1,s^{*}}) = p(\bc_{2,s^{*}}|\bx_{0,s^{*}}) \). Our studies demonstrate that this probability can be effectively represented by energy-based models: \( p(\bc_{2,s^{*}}|\bx_{0,s^{*}}) \propto \exp(-\gamma E(\bc_{2,s^{*}}, \bx_{0,s^{*}})) \), where \( E \) denotes a differentiable energy function and the partition function is neglected as after taking the logarithm, the partition function contributes only a constant term, which becomes zero if we take gradient on $\bx_{0,s^{*}}$. This energy function attains lower values for consistent pairs of simulation \( \bc_{2,s^{*}} \) and sample \( \bx_{0,s^{*}} \) and higher values for inconsistent pairs. For instance, if $\bc_{2,s^{*}}$ is a coarse-grained prediction for $\bx_{0,s^{*}}$, one could define $E$ as $E(\bc_{2,s^{*}},\bx_{0,s^{*}})=\norm{\bc_{2,s^{*}}-\bx_{0,s^{*}}}^2$.  \( \gamma \) serves as a temperature parameter that modulates the strength of the conditional probability. This model framework promotes consistency between the sample and the corresponding expensive physical simulation. We will explore two applications of the energy-based approach in Sections~\ref{sec:fluid}~and \ref{sec:heat}. 

\section{Theoretical Analysis}
\label{sec:theory}
Now, we provide theoretical guarantees for the sampling algorithms in the continuous regime. Remember that the raison d'etre for Algorithm~\ref{alg:sample} is to obtain samples from $p(\bx_{0,s^{*}}|\bx_{0,1:\sct},\bc_{1,s^{*}})$ and $p(\bx_{0,s^{*}}|\bx_{0,1:\sct},\bc_{1,s^{*}},\bc_{2,s^{*}})$. A natural performance metric for the sampler is thus the distance between the ground truth and sampling distribution. 

In this section, we use $\qcun(\bx_{0,s^{*}})$ to denote the distribution of samples generated by Algorithm~\ref{alg:sample} with choice 1, and $\qcdeux(\bx_{0,s^{*}})$ to denote the sample distributions from Algorithm~\ref{alg:sample} with choice 2, where we omit the dependence on $\bx_{0,1:\sct},\bc_{1,s^{*}}$, and $\bc_{2,s^{*}}$ for brevity. Then a well-behaving algorithm should satisfy $\qcun(\bx_{0,s^{*}})\approx p(\bx_{0,s^{*}}|\bx_{0,1:\sct},\bc_{1,s^{*}})$ and $\qcdeux(\bx_{0,s^{*}})\approx p(\bx_{0,s^{*}}|\bx_{0,1:\sct},\bc_{1,s^{*}},\bc_{2,s^{*}})$. Similar to~\cite{wassdiv}, we use the Wasserstain distance to quantify the difference between the sampling and ground truth distributions. The Wasserstein distance~\cite{otbook} between two p.d.f. $p_1$ and $p_2$ is denoted as $\wdiv{p_1(\bx)}{p_2(\bx)}$. 

Intuitively, the differences between the two distributions result from two sources. The first is the inaccurate denoising network: if $\veceps_{\theta}(\cdot)$ cannot learn the score function $\nabla_{\bx_{t,s^{*}}}\log p(\bx_{t,s^{*}}|\bx_{0,1:\sct},\bc_{1,s^{*}})$ to high precisions, the sampling distribution can be inaccurate. Mathematically, we define the expected $\ell_2$-error between the denoising network prediction and the ground truth score as,
\begin{align}
&\loss_1 = \frac{1}{2}\int_{0}^T\mathbb{E}_{\bx_{t,s^{*}}}\Big[\Big\lVert\frac{\veceps_{\theta}(\bx_{t,s^{*}},\bx_{0,1:\sct},\bc_{1,s^{*}})}{\sqrt{1-\bara_t}}\notag\\
&+\nabla_{\bx_{t,s^{*}}}\log p(\bx_{t,s^{*}}|\bx_{0:1\sct},\bc_{1,s^{*}})\Big\rVert^2\Big]\beta_t dt.
\end{align}

The second source of error originates from inaccurate $\bg$ functions: if $\bg$ does not accurately represent the gradient of the log conditional probability, the sampling algorithm can also be problematic. 
\begin{align}
&\loss_2 = \frac{1}{2}\int_{0}^T\beta_t\mathbb{E}_{\bx_{t,s}}\Big[\Big\lVert\bg(\bx_{t,s^{*}},\bx_{0,1:\sct},\bc_{1,s^{*}},\bc_{2,s^{*}})\notag\\
&-\nabla_{\bx_{t,s}}\log p(\bc_{2,s^{*}}|\bx_{t,s^{*}},\bx_{0,1:\sct},\bc_{1,s^{*}})\Big\rVert^2\Big] dt.
\end{align}

The following theorem provides an upper bound on the Wasserstein distance between the sampling and the ground truth distribution.
\begin{theorem}
\label{thm:wassdis}
Under regularity conditions, if we use Algorithm~\ref{alg:sample} with choice 1 to sample $\bx_{0,s^{*}}$, in the continuous limit, the sample distribution $\qcun$ satisfies,
\begin{align}
\label{eqn:kldivqch1}
&\wdiv{p(\bx_{0,s^{*}}|\bx_{0,1:\sct},\bc_{1,s^{*}})}{\qcun(\bx_{0,s^{*}})}\notag\\
&=O \left(\sqrt{\loss_1}+\wdiv{p(\bx_{T,s^{*}}|\bx_{0,1:\sct},\bc_{1,s^{*}})}{\mathcal{N}(0,\mathbf{I})}\right).
\end{align}

Similarly, if we use Algorithm~\ref{alg:sample} with choice 2 to sample $\bx_{0,s^{*}}$, in the continuous limit, the sampling distribution $\qcdeux$ would satisfy,
\begin{align}
\label{eqn:kldivqch2}
&\wdiv{p(\bx_{0,s^{*}}|\bx_{0,1:\sct},\bc_{1,s^{*}},\bc_{2,s^{*}})}{\qcdeux(\bx_{0,s^{*}})}=\notag\\
&O\left(\sqrt{\loss_1+\loss_2}+\wdiv{p(\bx_{T,s^{*}}|\bx_{0,1:\sct},\bc_{1,s^{*}},\bc_{2,s^{*}})}{\mathcal{N}(0,\mathbf{I})}\right).
\end{align}
\end{theorem}

It is worth noting that in practice, the forward diffusion processes are often designed carefully such that $p(\bx_{T,s}|\bx_{0,1:\sct},\bc_{1,s^{*}})$ and $p(\bx_{T,s}|\bx_{0,1:\sct},\bc_{1,s^{*}},\bc_{2,s^{*}})$ are extremely close to standard normal distributions~\cite{ddpm}. As such, their Wasserstein distance is often insignificant, and the right-hand side of~\eqref{eqn:kldivqch1} and~\eqref{eqn:kldivqch2} are dominated by $\sqrt{\loss_1}$ and $\sqrt{\loss_1+\loss_2}$.

There are a few implications from Theorem~\ref{thm:wassdis}. First,~\eqref{eqn:kldivqch1} indicates that if we use choice 1 from Algorithm~\ref{alg:sample}, the sampling distribution error is determined by the prediction error
of the denoising network $\loss_1$. This is consistent with our intuition that a more accurate denoising network $\veceps_{\theta}$ will lead to higher-quality samples. Second,~\eqref{eqn:kldivqch2} suggests that the sampling error for choice 2 is related to the estimation error in both denoising network $\veceps_{\theta}$ and the gradient of the log conditional probability model $\bg$. Accurate $\veceps_{\theta}$ and $\bg$ estimates would bring the distribution $\qcdeux$ close to the ground truth $p(\bx_{0,s^{*}}|\bx_{0,1:\sct},\bc_{1,s^{*}},\bc_{2,s^{*}})$. 

Inspired by~\cite{wassdiv
}, the proof of Theorem~\ref{thm:wassdis} follows from the contraction property of the Wasserstein distance. We relegate the complete proof to the supplementary materials.

\section{Fluid system}
\label{sec:fluid}
We first investigate the numerical performance of the proposed \name on a fluid system. The dynamics of viscous fluids are described by Navier-Stokes equations, which are nonlinear partial differential equations. In practice, Navier-Stokes equations are often solved by computational fluid dynamics (CFD) programs. Numerous CFD programs have been developed in recent decades, and many of them rely on finite difference methods that solve fluid fields on the grids~\cite{cfdbook}. Like many physics simulators, finite difference CFD methods face tradeoffs between grid resolution and simulation fidelity, which renders \name a useful tool to integrate the predictions from CFD simulations with different resolutions and leverage the combined knowledge to make predictions. \revise{This section presents the main experimental results, while additional details,  ablation studies, and two generated videos are provided in the supplementary
materials.}

\subsection{Experiment setup}
In the numerical study, we analyze the movement of 2D fumes driven by buoyancy and gravity, with the goal of predicting buoyancy fields. We use Boussinesq approximation~\cite{boussinesq} to analyze the evolution of the fume system. The ground truth data are generated by running multiple simulations of the fumes with existing high-performance CFD simulators~\cite{fluidsim,fluidfft} on fine-grained $128\times 128$ grids. More specifically, we randomly initialize the buoyancy and vorticity at time $s=0$ and run the simulator from $s=0$ to $s=10$. We use the buoyancy field at $s=0$ time steps as the context vectors, and predict the target state at $s^{*}=10$. \revise{Results of $N= 6880$ simulations from different random initializations are accumulated and then randomly separated into 90\% training set and 10\% test set.} We train the DeNN $\veceps_{\theta}$ on the training set. On the test set, we try to predict $\bx_{0,s^{*}}$ with the information $\bx_{0,1:\sct}$, $\bc_{1,s^{*}}$, and possible $\bc_{2,s^{*}}$. The buoyancy at $s^{*}=10$ for four samples from the test set is plotted in the last column of Fig.~\ref{fig:fumes}.  

To apply \name, inexpensive physics predictions $\bc_{1,s^{*}}$ and expensive physics predictions $\bc_{2,s^{*}}$ are needed. We generate these predictions using the same Navier-Stokes-Boussinesq simulator but on coarser grids. More specifically, for each simulation, we run the fluid simulator from the same initialization as the ground truth but with a grid resolution of $32\times 32$. The buoyancy and vorticity fields at $s^{*}=10$ are the inexpensive physics prediction $\bc_{1,s^{*}}$. Similarly, we run the fluid simulator with a grid resolution of $64\times 64$ and use the buoyancy as $\bc_{2,s^{*}}$. 

Four samples of $\bc_{1,s^{*}}$ are plotted in the first column of Fig.~\ref{fig:fumes}. One can observe that the inexpensive simulation can capture the low-frequency patterns of the ground truth while details of fume swirls are blurred. This disparity demonstrates the inherent simulation bias. 

In this study, we choose the conditional probability model $\log p(\bc_{2,s^{*}}|\bx_{0,s},\bc_{1,s^{*}})$ as,
\begin{align}
\label{eqn:patchmodel}
&\log p(\bc_{2,s^{*}}|\bx_{0,s},\bc_{1,s^{*}}) =C+\notag\\
&-\gamma \norm{\text{AvgPool}^{2\times 2}(\bc_{2,s^{*}})-\text{AvgPool}^{4\times 4}(\bx_{0,s^{*}})}^2, 
\end{align}
where $\text{AvgPool}^{2\times 2}$ is an average pooling operation~\cite{resnet} for the patch size of $2$ by $2$. More precisely, it divides the $64\times 64$ buoyancy field into $32\times 32$ patches of size $2\times 2$, then calculates the average buoyancy in each patch. Similarly, $\text{AvgPool}^{4\times 4}$ is a $4\times 4$ pooling operation. The model~\eqref{eqn:patchmodel} encourages the low-frequency information of the simulated buoyancy and predicted buoyancy to be matched. $\gamma$ is a coefficient that measures the confidence of the result from the simulated buoyancy. We simply set it to be $0.01$ throughout our experiments. $C$ is a normalization constant that will become zero after differentiation. Additionally, we find removing the coefficient $1-\alpha_t$ in~\eqref{eqn:term3} of~\eqref{eqn:expensivediffusionsample} and directly using $\bg$ in the sampling update is more numerically stable. Therefore, we implement this modified version of the sampling update.

\subsection{Benchmarks and visual results}
\label{sec:fumebenchmark}
With access to physics simulations, we can implement \name to predict the ground-truth buoyancy. We use a U-Net~\cite{unet} implemented in~\cite{unetimplement} as the DeNN. The details of U-net architecture are elaborated in the
supplementary materials.

\begin{figure*}
    \centering
    \includegraphics[width=0.9\textwidth]{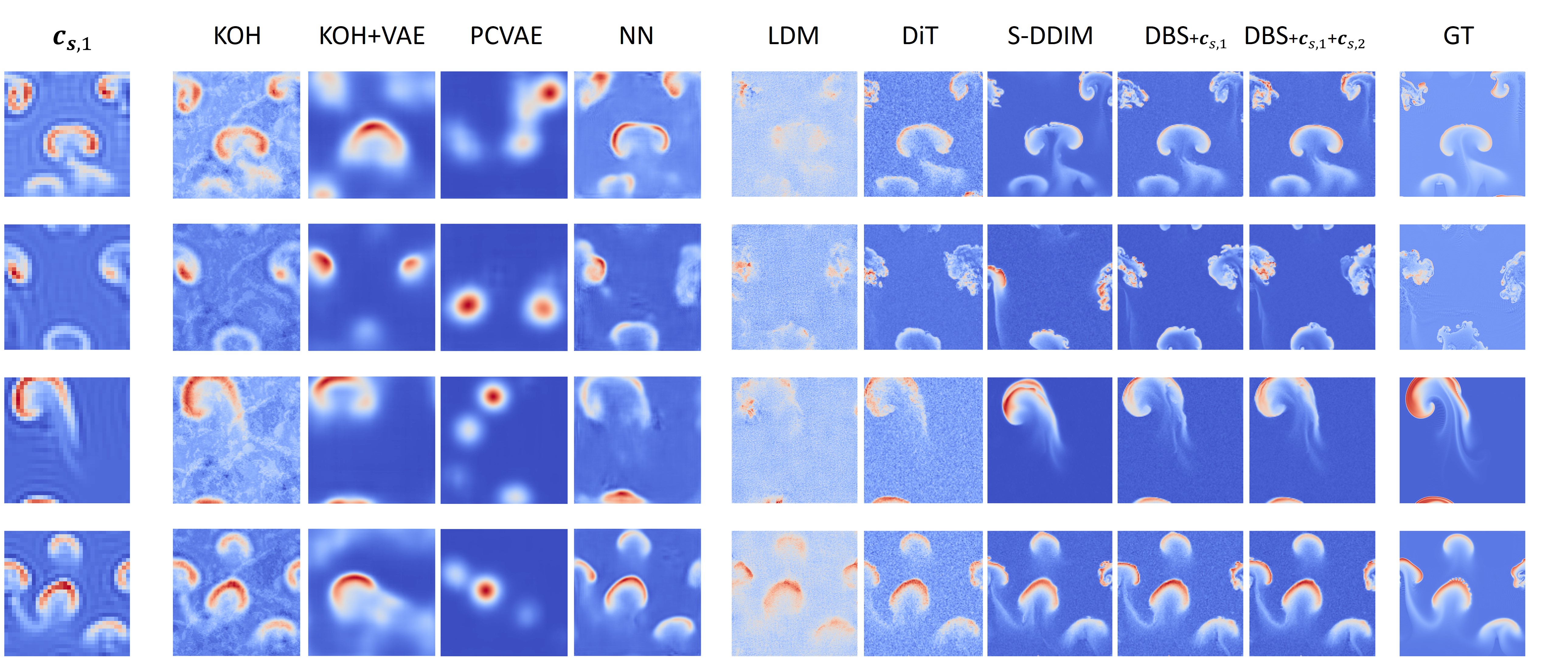}
    \vspace{-2mm}
    \caption{\revise{Illustrations of the ground truth bouyancy, physics predictions $\bc_{s,1}$, and predictions from $5$ different models. $4$ random samples are plotted from the test set. Red denotes large buoyancy and blue denotes low buoyancy. The ``$\bc_{s,1}$'' column contains the output from inexpensive simulations. ``DDIM'' represents samples generated by standard DDIM without any physics conditioning. ``\name+ $\bc_{s,1}$'' represents the samples from Algorithm~\ref{alg:sample} with choice 1. ``\name+ $\bc_{s,1}+\bc_{s,2}$'' represents samples from Algorithm~\ref{alg:sample} with choice 2.} }
    \label{fig:fumes}
\end{figure*}

\revise{For comparison, we implement several benchmark algorithms that represent the state-of-the-art in Bayesian calibration and diffusion generative models.
\begin{itemize}
    \item KOH~\cite{kennedy2000predicting}: The Bayesian calibration algorithm (KOH) uses a GP to model the residuals between the physics simulation output and the ground truth. We implement KOH to predict the difference between the ground truth buoyancy and the buoyancy from simulation using a batch-independent multi-output GP model provided by GPytorch~\cite{gpytorch}.
    \item KOH with variational auto-encoder~\cite{kingma2019introduction}: we train a variational auto-encoder~\cite{diffusevae}, then implement the standard KOH on the latent space.
    \item Physics constrained variational auto-encoders (PCVAE)~\cite{pcvae}: we jointly train a variational auto-encoder and latent space dynamical model as described in~\cite{pcvae}.    
    \item NN~\cite{pinn}: We directly train a deep neural network (a U-Net~\cite{unet}) to predict the future states of a system given historical information. The network is trained without inputs of $\bc_{1,s^{*}}$ and $\bc_{2,s^{*}}$.
    \item Standard diffusion (S-DDIM): We implement the DDIM method described in Section~\ref{sec:standarddiff} to sample target state vectors without the information from physics simulations.
\item DiT~\cite{dit}: we implement \name by training a DiT-B-4 model from random initializations to learn the score functions of the buoyancy field with contextual input $\bc_{s,1}$.
    \item Latent diffusion model~\cite{ldm}: we use a pre-trained auto-encoder~\cite{diffusers} to transform high-resolution buoyancy field into low-dimensional latent features, then implement Algorithm~\ref{alg:trainepsilon} in the latent space with contextual input $\bc_{s,1}$.
\end{itemize}}
 The predictions of benchmark algorithms and \name are plotted in Fig.~\ref{fig:fumes}.

\revise{From Fig.~\ref{fig:fumes}, we can clearly see that S-DDIM predictions exhibit sharp and vivid details on the small-scale swirling structures of the fume. However, locations of large-scale swirl patterns are not accurate. This indicates that the DeNN learns the localized buoyancy patterns effectively but struggles to understand the long-range physical knowledge. The limitation is addressed when we use inexpensive physics prediction $\bc_{1,s^{*}}$ as an additional input to the denoising network. Conditioned on the inexpensive physics prediction, samples from \name align more closely with the ground truth in terms of large-scale structures. The results imply that the physics knowledge is integrated in \name, which reaps benefits from the simulation while mitigating its bias. Furthermore, when $\bc_{2,s^{*}}$ is available and used in choice 2 of Algorithm~\ref{alg:sample}, the fidelity of the buoyancy prediction further improves, indicating that the more refined physics simulations can boost the quality of prediction. }

\revise{In Fig.~\ref{fig:fumes}, DiT-generated samples often contain noisy backgrounds. This is likely due to the relatively small size of our training set (6192 samples), whereas state-of-the-art DiT models~\cite{dit} are trained on large-scale datasets such as ImageNet~\cite{imagenet} with 1.2 million images. Given that transformer-based architectures lack the inductive bias of spatial smoothness, they may not perform well in low-data regimes. LDM does not generate high-quality predictions either. We hypothesize that this is due to domain shifts: the encoders used in LDM are pre-trained on standard image datasets, whereas fluid buoyancy fields exhibit different patterns. These domain shifts lead to a complicated distribution of latent features, which DDMs struggle to learn well. } 

\revise{For non-diffusion models, the standard KOH does not add meaningful information to the physics simulation, probably because the Gaussian process is not expressive enough in the high-dimension regime. As our prediction is a $128\times 128=16382$ dimensional vector, predicting it using a Gaussian process is uneasy. KOH+VAE generates samples with vague edges, suggesting that the fine details are not well captured. The PCVAE approach does not give accurate predictions either. This is understandable as approximating the dynamics of high-dimensional buoyancy fields by evolutions of low-dimensional latent features is challenging. The NN approach also generates vague samples, probably because it completely relies on the training set to learn the physical evolutions of the fume system. Such a purely statistical approach may not produce decent performance when the training set is not extremely large.}

\subsection{Numerical performance}
\label{sec:fumenumerical}
For numerical comparisons, we also evaluate the quality of the prediction on four standard evaluation metrics: mean squared error (MSE), peak signal-to-noise ratio (PSNR), structural similarity index (SSIM), and learned perceptual image patch similarity (LPIPS)~\cite{lpips}. In general, a lower MSE, a higher PSNR, a higher SSIM, and a lower LPIPS indicate a better sample quality. Amongst these metrics, MSE and PSNR measure the low-level pixel-wise difference between the sample and the ground truth, while SSIM is more aligned with human perception of the images. LPIPS leverages a deep neural network (VGG) to capture high-level visual similarities. The mean and standard deviation on the test set are reported in Table~\ref{tab:fumes}. In each column, the best result is highlighted in \textbf{bold}, while the second-best result is emphasized with an \underline{underline}.
\begin{table}[htbp]
  \centering
  \caption{\revise{The mean and standard deviation of the sample quality of different algorithms.}}
\revise{\begin{tabular}{ccccc}
    \toprule
          & MSE (0.001)$\downarrow$   & PSNR$\uparrow$  & SSIM $\uparrow$ & LPIPS$\downarrow$ \\
    \midrule
    KOH   & $1.33(0.04)$ & 29.8(0.1) & 
    99.986(0.001) & 0.416(0.003) \\
    KOH{\small+VAE}   & $1.58(0.07)$ & 29.4(0.2) & 
    99.982(0.001) & 0.301(0.007) \\
    PCVAE  & 2.66(0.12) & 26.7(0.2) & 99.966(0.001) & 0.338(0.004) \\
    NN  & 1.33(0.06) & 31.2(0.3) & 99.987(0.001) & 0.292(0.008) \\
    \midrule
    S-DDIM & 3.69(0.2) & 25.5(0.2) & 99.955(0.001) & 0.401(0.002) \\
    LDM & 2.42(0.05) & 26.6(0.1) & 99.984(0.001) & 0.335(0.006) \\
    DiT & 1.50(0.07) & 30.5(0.2) & 99.990(0.001) & 0.456(0.005) \\
    With $\bc_{s,1}$  & \underline{1.14}(0.08) & \underline{33.2}(0.4) & \underline{99.992}(0.001) & \underline{0.287}(0.009) \\
    With $\bc_{s,2}$  & \textbf{1.00}(0.05) & \textbf{33.6}(0.3) & \textbf{99.993}(0.001) & \textbf{0.228}(0.006) \\
    \bottomrule
    \end{tabular}}
  \label{tab:fumes}%
\end{table}%
Results presented in Table~\ref{tab:fumes} corroborate visual observations depicted in Fig.~\ref{fig:fumes}. S-DDIM is capable of generating samples that visually resemble the ground truth, as evidenced by its low LPIPS scores when compared with those from KOH and NN methods. However, the model's high MSE and low PSNR indicate a poorer alignment with the ground truth at pixel levels.

With inexpensive physics predictions $\bc_{1,s^{*}}$, \name enhances the sample quality as the LPIPS decreases. Furthermore, there are noticeable improvements in MSE, PSNR, and SSIM relative to the standard DDMs. Incorporation of the expensive physics simulation $\bc_{2,s^{*}}$ into the reverse diffusion process~\eqref{eqn:expensivediffusionsample} leads to even greater improvements. The LPIPS scores decrease further, and the MSE, PSNR, and SSIM reach the highest values compared to all other evaluated algorithms. These improvements substantiate the benefits of utilizing multiple simulations in Algorithm~\ref{alg:sample}.

\revise{In comparison, the benchmark KOH, KOH+VAE, PCAVAE, and NN incur high LPIPS, corroborating our observations that these samples do not show consistent visual patterns as the ground truth. The comparisons highlight the advantages of the proposed \name in producing high-quality results.}

\section{Thermal process in 3D printing}
\label{sec:heat}
We further apply \name on a real-life process of laser-based metal additive manufacturing (LBMAM). LBMAM uses a laser beam to heat and melt metal powders deposited on the printbed to print 3D objects layer by layer~\cite{1,pfe}. To monitor the manufacturing process, a thermal camera is installed to capture in-situ temperature distribution on the printing surface. The thermography provides rich information for characterizing the process and potentially identifying defects~\cite{40,41}. From a statistics perspective, we aim to predict a future frame of the thermal video, given some observed frames and the information about the movement of the laser that can be recovered from the G-code of the printer. 

\begin{figure}[h]
    \centering
    \includegraphics[width=0.25\textwidth]{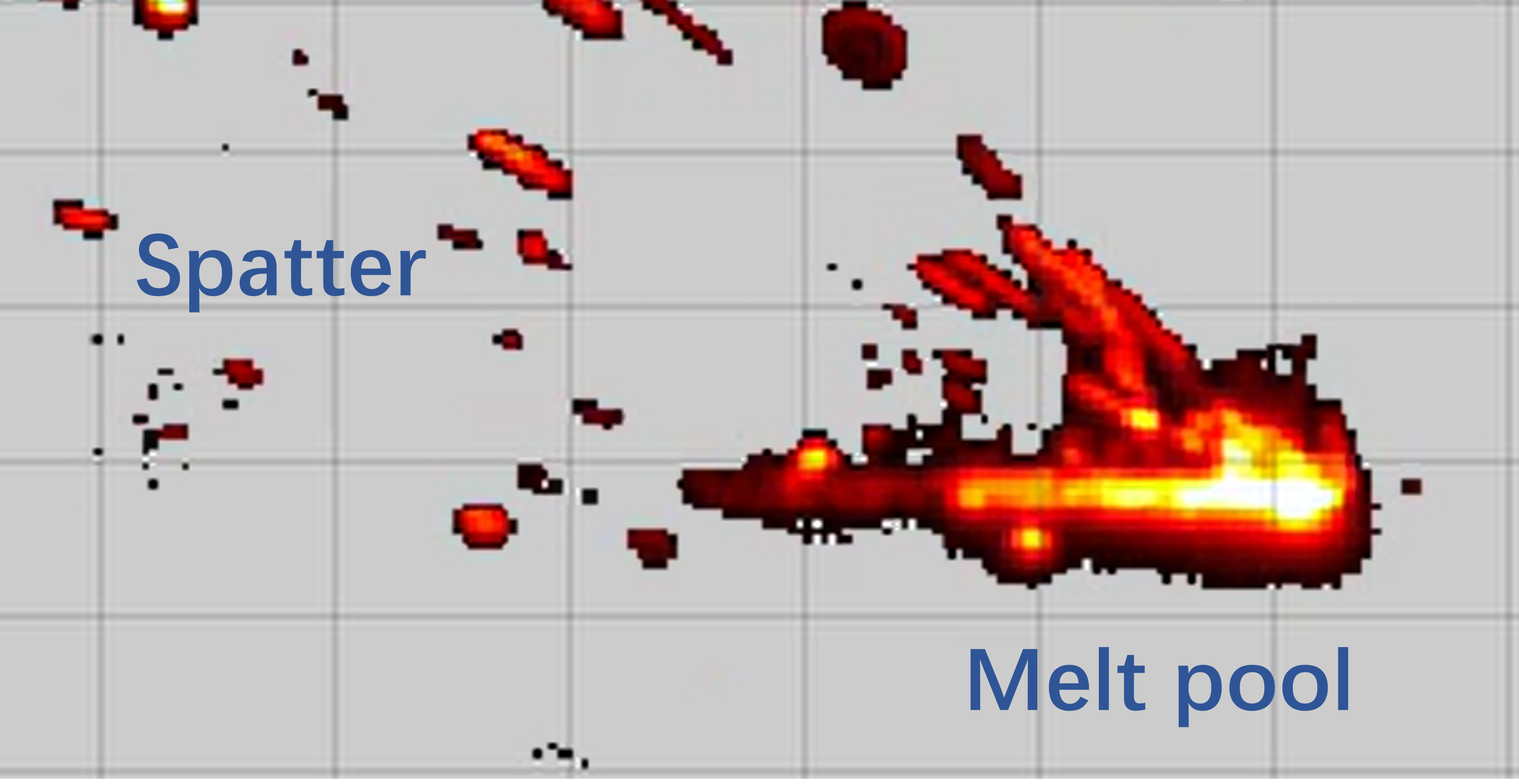}
    \vspace{-2mm}
    \caption{The comet-shaped melt pool and spatters in LBMAM.}
    \label{fig:lbmam}
\end{figure}

A typical thermal frame from an LBMAM process is plotted in Fig.~\ref{fig:lbmam}, where the underlying physics in the thermal process can roughly be divided into the melt pool's heat dissipation and the spatters' movement~\cite{42,6}. We use two ad-hoc physics models to simulate them. The melt pool is the connected high-temperature region around the laser beam. Its dynamics are described by a 2D heat equation, which has a simple closed-form solution in the ideal case. The solution is easy to calculate; hence we model it as $\bc_{1,s^{*}}$. The spatters are more volatile and irregular. We use a flow model to describe its movement. Compared to the heat pool PDE, the flow model incurs a higher computational cost. Thus, we model the flow velocity field as $\bc_{2,s^{*}}$. We will describe the details of the two models in this section. It is worth noting that both physics models are oversimplified and thus biased, yet our goal is to fuse data with inaccurate physics simulations to get superior predictions.

For notational consistency, we use $\bx$ to denote the pixeled vectorized frame in a thermal video. If the frame has resolution $W\times H$, then $\bx$ is a vector $\bx \in \mathbb{R}^{WH}$. With a slight abuse of notation, we use $(x,y)$ to denote the continuous spatial coordinate on the 2D plane. We sometimes use the abbreviated notation $\br=(x,y)^{\top}$. Furthermore, we use $u(x,y,s)$ to denote a time-varying temperature field on 2D: $u:\mathbb{R}^2\times\mathbb{R}^{+}\to \mathbb{R}$. We use subscripts $u_m$ and $u_p$ to represent the temperature field of the melt pool and spatters. A thermal video frame $\bx$ naturally corresponds to the temperature $u$ at $W$ by $H$ grid locations.

\subsection{Dynamics of the melt pool}
The melt pool often displays comet-like shapes, which is a result of heat dissipation and laser movement. In this section, we will construct a computationally amenable model to simulate the morphology and dynamics of the melt pool.

\subsubsection{Heat equation}
We use $u_m(x,y,s)$ to denote the melt pool temperature at point $(x,y)$ at time $s$. 

2D heat equation naturally models the physics of $u_m$,

\begin{equation}
\label{eqn:2dheatequation}
    \frac{\partial u_m}{\partial s} = \nabla\cdot \kappa \nabla u_m  - \rho u_m +f(x,y,s),
\end{equation}
where the term $\nabla\cdot \kappa \nabla u_m $ represents heat dissipation. $\kappa$ is the thermal diffusivity matrix, $
\kappa = \left(
\begin{aligned}
\kappa_x &\quad 0\\
0 &\quad \kappa_y
\end{aligned}
\right).
$

Term $- \rho u_m$ represents the loss of heat from the printing surface into the air, and $f(x,y,s)$ represents the energy injected by the laser beam at $(x,y,s)$.

\subsubsection{Solution in the ideal case}
\label{sec:heatequationsolution}

Exactly solving \eqref{eqn:2dheatequation} is uneasy as the diffusivity $\kappa$ can be anisotropic and temperature-dependent and location-dependent, and the same for the parameter $\rho$. The boundary condition can also be complicated. Conventional physics simulators are often based on discrete element analysis~\cite{ampde} or SPH~\cite{amsph}. 

To obtain a conceptually simple and computationally tractable physical solution, we can make a few simplifying assumptions. We assume $\kappa$ and $\rho$ are temperature-independent constants. The initial condition is $u_m(x,y,0) = 0$. And the 2D plane extends to infinity.

Then the solution to \eqref{eqn:2dheatequation} is given by,
\begin{align}
\label{eqn:heatsolution}
&u_m(x,y,s;\phi) = \int_{-\infty}^{\infty}\int_{-\infty}^{\infty} \int_{\spr=0}^{s} G(x,y,s;\xp,\yp,\spr;\phi)\notag\\
&\times f(\xp,\yp,s^{\prime})ds^{\prime}d\xp d\yp,
\end{align}
where $G(x,y,s;\xp,\yp,\spr;\phi)$ is the Green's function,
\begin{align}
\label{eqn:greensfunction}
&G(x,y,s;\xp,\yp,\spr;\phi) = C_n\exp\left(-\rho(s - s^{\prime})\right)\notag\\
&\times\exp\left(-\frac{(\br-\brp)^{\top}\kappa^{-1}(\br-\brp)}{4(s - s^{\prime})}\right),
\end{align}
and $\phi=(\rho,\kappa_x,\kappa_y,C_n)$ denotes the system parameters. 

In \eqref{eqn:heatsolution}, $f(x,y,s)$ is given by the Gcode of 3D printers. Therefore, the evolution of temperature $u_m$ is controlled by only $4$ parameters $\kappa_x$, $\kappa_y$, $\rho$, and $C_n$. We calibrate these four parameters from data using nonlinear least squares. More specifically, on a dataset of observed temperature on grid $\{u_m^{(i)}\}_{i=1}^N$, we optimize parameter $\phi$ to fit the observations by empirical error minimization $
\min_{\phi}\frac{1}{N}\sum_{i=1}^N\norm{u_m^{(i)}-u_m(\cdot;\phi)}^2$, where $u_m(\cdot;\phi)$ denotes the temperature $u_m$ evaluated at grid points. Since the empirical loss is differentiable, the minimization is implemented by Adam. Though $C_n$ is fixed in theoretical derivation~\eqref{eqn:greensfunction}, we still calibrate it from data to adapt to the different scalings of $f$ and $u_m$. With calibrated parameters $\hat{\phi}$, we use the value of $u_m(\cdot;\hat{\phi})$ from equation~\eqref{eqn:heatsolution} at $W$ by $H$ grid points as $\bc_{1,s^{*}}$ in Algorithm~\ref{alg:sample}.

\subsection{Dynamics of the spatters}
\label{sec:spatterdynamics}
The interaction between high-energy laser beams and metal powders generates high-temperature particles that scatter from the metal surface. These particles can also be captured by the thermal camera. However, the 2D heat dissipation PDE can hardly describe the movement of particles. As a result, an alternative physics model is needed.

As the high-temperature particles are heated by the energy from the laser, it is natural to model them to be created at the center of the meltpool. After generation, these particles will move toward the edge of the receptive field of the thermal camera. The optical flow model~\cite{opticalflow} provides a suitable characterization for such an emanating movement pattern. 

More specifically, we define a velocity field $\bv(x,y,s)=[v_x,v_y]^{\top}\in\mathbb{R}^2$ denoting the velocity of the particle at position $(x,y)$ and time $s$. The velocity field $\bv$ is the expensive physics simulation $\bc_{2,s}$. We use $v_x$ to denote the velocity along the $x$-axis and $v_y$ to denote the velocity at the $y$-axis. For a small time interval $\Delta s$, the temperature of the scattering particles should remain approximately constant. As a result, $u_p$ should satisfy,
\begin{align*}
u_p(x,y,s)=u_p(x-v_x\Delta s,y-v_y\Delta s,s-\Delta s).
\end{align*}
This equation is widely employed in the literature of optical flows~\cite{opticalflow}. Based on the rationale, we design the following probability model to characterize the movement of spatters,
\begin{align}
\label{eqn:flowphysicsprobability}
&\log p(\bv|\bx_{0,s^{*}},\bx_{0,s^{*}-1})=C+\notag\\
&-\frac{\gamma}{2}\norm{\mathcal{P}_{\Omega_\text{spatter}}\left(\wrap\left(\bx_{0,s^{*}-1},\bv\right)-\bx_{0,s^{*}}\right)}^2,
\end{align}
where $\bv$ plays the role of the expensive simulation $\bc_{2,s^{*}}$.

In~\eqref{eqn:flowphysicsprobability}, $\wrap$ denotes the warping operator, which employs a semi-Lagrangian method to infer $\bx_{0,s^{*}}$ from $\bx_{0,s^{*}-1}$ and $\bv$. A more detailed definition of $\wrap$ will be provided in the supplementary material. $\mathcal{P}_{\Omega_{\text{spatter}}}$ denotes the projection into the spatter region, $\Omega_\text{spatter}$. More specifically,  $
[\mathcal{P}_{\Omega_{\text{spatter}}}(\bx)]_{j}=
    [\bx]_{j}$ if $j$ is a grid point in $\Omega_{\text{spatter}}$ and $[\mathcal{P}_{\Omega_{\text{spatter}}}(\bx)]_{j}=0$ otherwise. 
Intuitively, $\mathcal{P}_{\Omega_{\text{spatter}}}$ applies a mask that selects the spatter temperature distribution from the entire temperature field. In our implementation, we define $\Omega_{\text{spatter}}$ as the complement of the melt pool region, which we estimate from the PDE solution~\eqref{eqn:heatsolution}. 

It is important to highlight that both $\wrap$ and $\mathcal{P}_{\Omega_{\text{spatter}}}$ are differentiable. 
Hence, we can plug in~\eqref{eqn:flowphysicsprobability} into~\eqref{eqn:gdef} to calculate $\bg$ using auto-differentiation packages. Subsequently, the gradient $\bg$ is used in Algorithm~\ref{alg:sample} to guide the sampling process with the information from flow field $\bv$. 


\subsection{Case study}
We use a subset of the 2018 AM Benchmark Test Series from NIST~\cite{nistdataset} as a testbed of the LBMAM thermal video prediction model. The dataset is collected in an alloy LBMAM process of a bridge structure manufactured in $624$ layers. An infrared camera with a frame rate of $1800$ frames per second is installed for in-situ thermography. 

We select the thermal videos captured when printing the first $50$ layers and slice them into video clips, each of which contains $10$ frames. The clips where the laser beam moves outside the camera receptive field are discarded since they are not informative. Each clip is a trajectory of the thermal process of LBMAM. On each clip, we use the beginning $2$ frames as the context frames and predict the last $5$ frames. We also perform an 80\%-20\% train-test splitting and present the results on the test set.

\begin{figure}[h]
    \centering
    \includegraphics[width=0.5\textwidth]{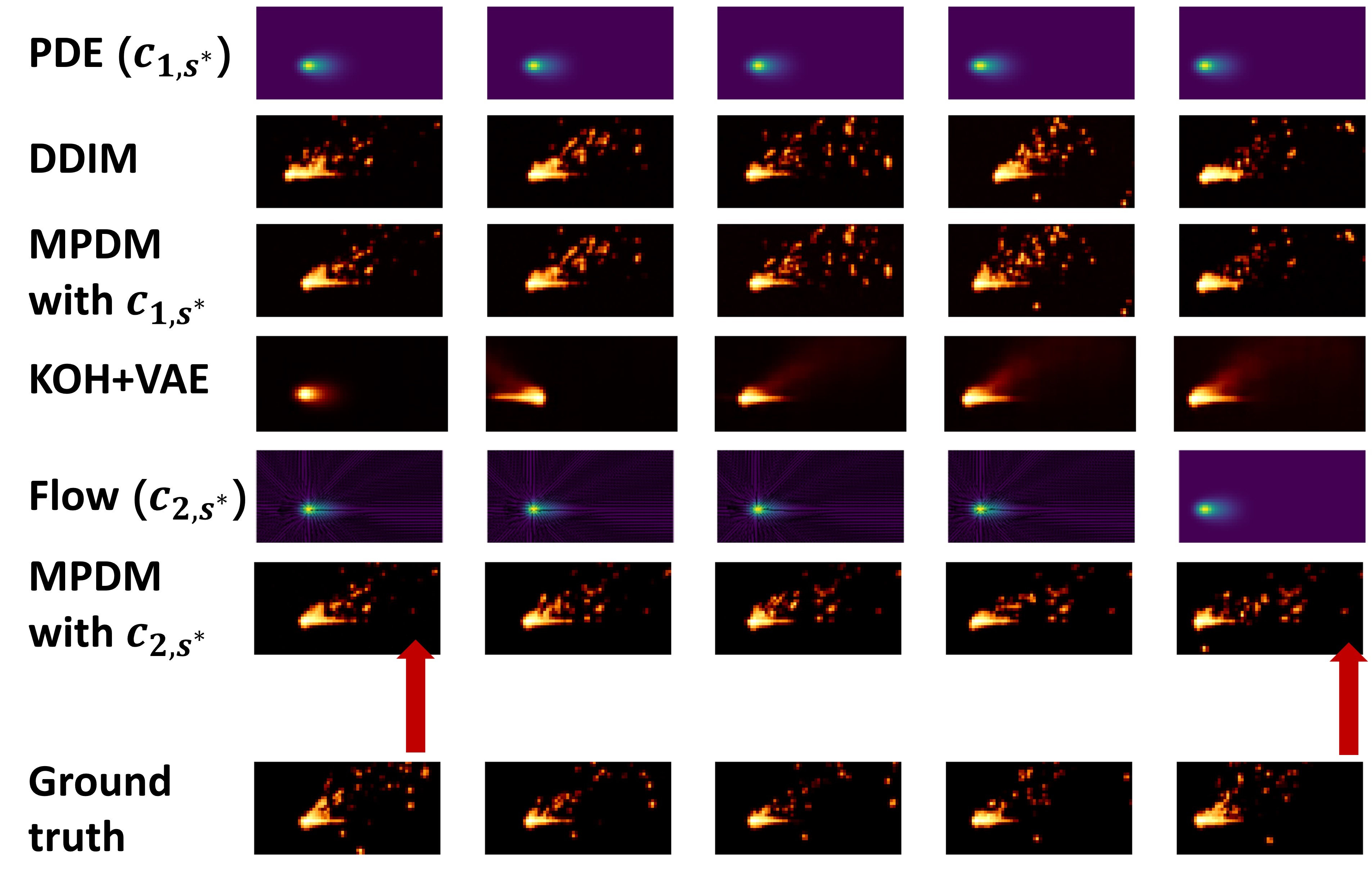}
    \vspace{-2mm}
    \caption{\revise{\name with guidance from PDE (inexpensive physics) and flow information (expensive physics) on five test frames. The directions of spatter flow are denoted as black arrows on the 4-th row. White denotes high temperature, and black denotes low temperature.}}
    \label{fig:double_physics}
\end{figure}

As discussed, the solution~\eqref{eqn:heatsolution} at $s^{*}$ is modeled as inexpensive physics predictions $\bc_{1,s^{*}}$, which is plotted as the first row of Fig.~\ref{fig:double_physics}. PDE solutions precisely identify the location of the melt pool. However, compared to the real thermal frames, the PDE solutions are excessively smooth, failing to capture the complex geometries of the melt pool and the spatters.

Conversely, S-DDIM generates more realistic and intricate temperature distributions. As illustrated in the second row of Fig.~\ref{fig:double_physics}, the irregular contours of the melt pool and the spatial arrangement of spatters bear a closer resemblance to those observed in ground truth data. Nevertheless, the positioning of the melt pool, as predicted by S-DDIM, is inaccurate, indicating deficiencies in the model's capability to track the dynamics of melt pool movement.

\name integrates the strengths of both PDE-based solutions and the diffusion model. Using the available $\bc_{1,s^{*}}$, we train the DeNN by Algorithm~\ref{alg:trainepsilon}, as elaborated in Section~\ref{sec:physicsguidedmodel}. The results are shown in the third row of Fig.~\ref{fig:double_physics}. These results vividly illustrate how DDMs can enhance the fine geometric details of the melt pool on top of PDE solutions. Our approach improves the accuracy of predicting the locations of the melt pool while simultaneously preserving the detailed realism of the temperature field.

\revise{Though single-frame predictions of inexpensive physics-informed diffusion demonstrate verisimilitude, the spatter patterns across frames are not consistent in Fig.~\ref{fig:double_physics}. Therefore, we add flow information to model the evolution of spatters as described in Section~\ref{sec:spatterdynamics}. The estimated flow fields on the test frames are plotted as small black arrows in the fourth row of Fig.~\ref{fig:double_physics}. Then, we use the flow field $\bv$ as the expensive physics simulation $\bc_{2,s^{*}}$ and sample thermal frames with Algorithm~\ref{alg:sample} with choice 2. It is worth noting that we set the spatial resolution of flow $\bv$ to be consistent with $\bx$. Results are plotted in the fifth row of Fig.~\ref{fig:double_physics}. As highlighted by the red arrows, the flow information makes spatter movement patterns more consistent.}

\begin{figure}[h]
    \centering
    \includegraphics[width=0.4\textwidth]{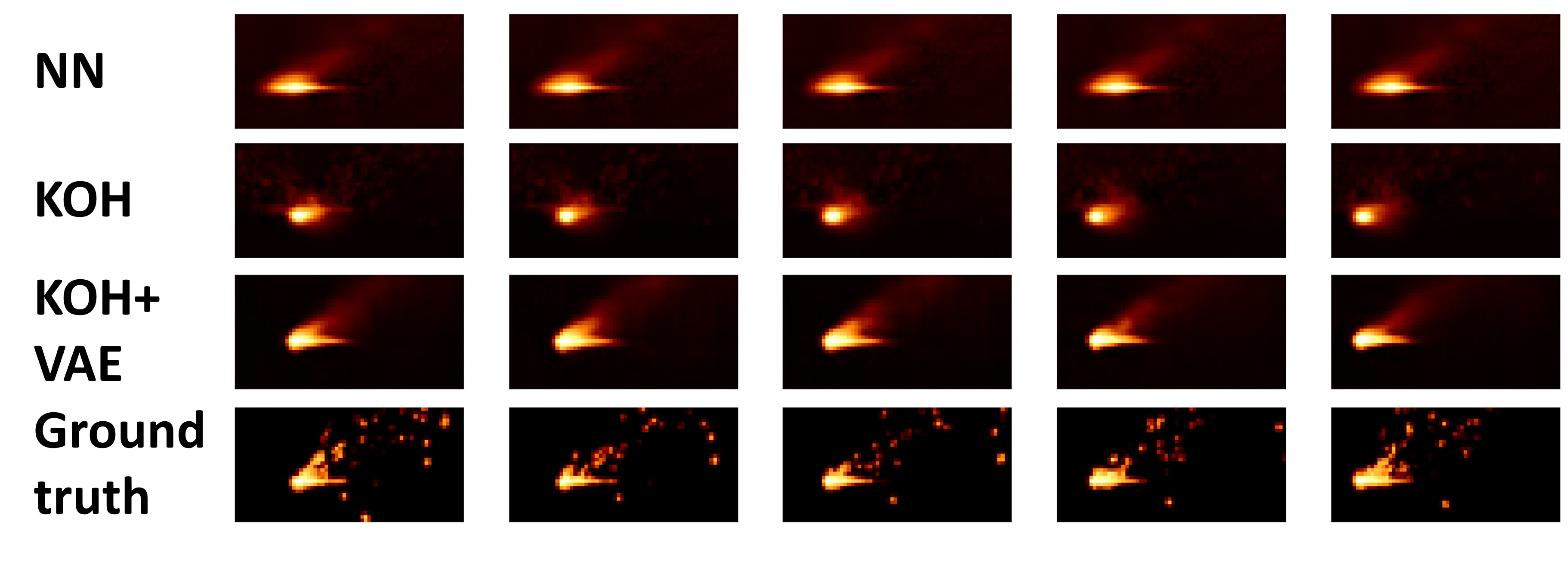}
    \vspace{-2mm}
    \caption{\revise{Frame predictions by benchmark methods NN, KOH, and KOH+VAE}.}
    \label{fig:heat_benchmark}
\end{figure}

\revise{For comparisons, we also implement KOH, KOH+VAE, and NN as benchmarks. Detailed implementation procedures conform to the specifications described in Section~\ref{sec:fumebenchmark}. The resulting predictions are illustrated in Fig.~\ref{fig:heat_benchmark}.}

\revise{In Fig.~\ref{fig:heat_benchmark}, results from all benchmarks are visually distinct from the ground truth. Specifically, the temperature distributions produced by NN appear over-dispersed. Similarly, the predictions by KOH lack meaningful spatter patterns observed in the ground truth data. Although the
KOH+VAE approach accurately predicts the meltpool’s shape
and location, it struggles to reproduce any spatters. When compared to the results depicted in Fig.~\ref{fig:double_physics}, it is evident that both PINN and KOH yield predictions with geometric patterns that are less consistent with those of the ground truth.}

\subsection{Numerical evaluation}
To thoroughly assess the performance of different approaches, we also calculate the PSNR, SSIM, and LPIPS between the predictions and the ground truth. The detailed settings are similar to Section~\ref{sec:fumenumerical}. Additionally, to evaluate the cross-frame consistency, we leverage the flow model to evaluate the consistency score between frames. The consistency score is defined as
\begin{equation*}
\text{Consistency Score}=\sum_{s}\frac{\norm{\mathcal{P}_{\Omega_{\text{spatter}}}(\wrap\left(\bx_{0,s-1},\bv_s\right)-\bx_{0,s})}^2}{\norm{\mathcal{P}_{\Omega_{\text{spatter}}}(\bx_{0,s-1}) }},
\end{equation*}
where $\bv_s$ is the flow prediction at time $s$. The consistency score measures how well the movement of spatters in samples complies with the flow model. Apparently, a lower consistency score signifies a higher level of cross-frame realism. The mean and standard deviation of different metrics evaluated on the test set are reported in Table~\ref{tab:performancecomparison}.

\begin{table}[ht]
\caption{\revise{Means and standard deviations of the evaluation metrics for different algorithms. CS denotes the consistency score.}}
\label{tab:performancecomparison}
\centering
\scalebox{0.99}{
    \begin{tabular}{ccccc}
    \toprule
     & PSNR$\uparrow$  & SSIM$\uparrow$  & LPIPS$\downarrow$ & CS$\downarrow$ \\
        \midrule
    KOH   & \textbf{25.1}(0.4) & \underline{99.976}(0.003) & 0.33(0.02) & \underline{0.28}(0.03) \\
    KOH +VAE  & 20.7(0.6) & 99.899(0.003) & 0.27(0.01) & \underline{0.28}(0.02) \\
    NN   & 24.0(0.2) & 99.960(0.002) & 0.24(0.01) & \textbf{0.064}(0.008) \\
    \midrule
    S-DDIM  & 21.3(0.2) & 99.924(0.004) & 0.166(0.002) & 1.43(0.03) \\
    With $\bc_{s,1}$    & \underline{24.3}(0.2) & \textbf{99.977}(0.006) & \underline{0.140}(0.002) & 1.41(0.02) \\
    With $\bc_{s,2}$   & 24.2(0.2) & \underline{99.976}(0.007) & \textbf{0.136}(0.005) & \textbf{0.064}(0.005) \\
    \bottomrule
    \end{tabular}

 }
\end{table}

Table~\ref{tab:performancecomparison} shows that the PDE solution $\bc_{1,s^{*}}$ significantly improves the sample quality for the diffusion model, as suggested by increases in PSNR and SSIM and a decrease in LPIPS. Such results are consistent with the visual observations in Fig.~\ref{fig:double_physics}. Additionally, the use of flow fields $\bc_{2,s^{*}}$ further improves the sample quality and drastically decreases the consistency score, which validates the observation in Fig.~\ref{fig:double_physics} that the spatter patterns move more consistently. 

\revise{For benchmark methods, though KOH attains a high PSNR value, its LPIPS score is high, suggesting that while KOH's predictions are accurate at a pixel level, they do not align well with human perceptual judgments. This observation is supported by Fig.~\ref{fig:heat_benchmark}, which shows that the modifications introduced by the KOH model to the PDE solutions are minor and barely perceptible. Therefore, similar to PDE solutions, KOH predictions lack spatters. KOH+VAE could not predict spatters either. Such results differ significantly from the ground truths in geometric patterns, thus incurring high LPIPS values. Results for the NN model suggest that its sample quality is not satisfactory either compared to \name with choice 2.}

\subsection{Uncertainty quantification}
\begin{figure}[h]
    \centering
    \includegraphics[width=0.4\textwidth]{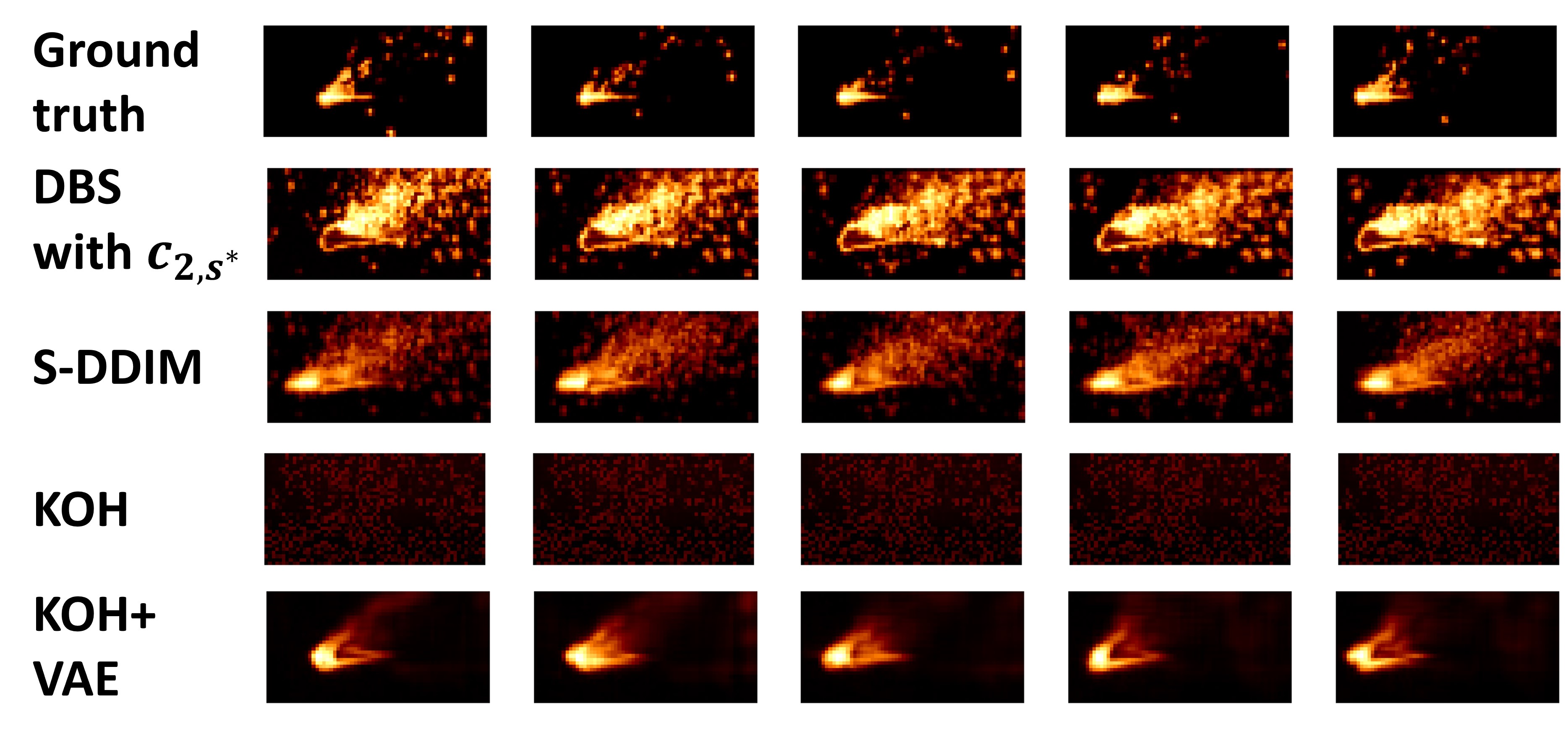}
    \vspace{-2mm}
    \caption{\revise{The predictive uncertainty for different methods. S-DDIM stands for standard diffusion without any physics input. White values denote higher levels of uncertainty, and black values denote lower levels of uncertainty.}}
    \label{fig:heat_uq}
\end{figure}

The probabilistic nature of \name provides a straightforward approach to uncertainty quantification. In this case study, we independently sample $40$ samples according to standard DDMs and Algorithm~\ref{alg:sample} with choice 2, and calculate pixel-wise standard deviations of the $40$ samples. The results are shown in Fig.~\ref{fig:heat_uq}.

Among benchmark algorithms, only KOH is capable of uncertainty quantification. We thus also plot the standard deviation of the learned GP in KOH in Fig.~\ref{fig:heat_uq}.

\revise{Fig.~\ref{fig:heat_uq} illustrates that the predictive variance of \name is high in spatter regions and low within the melt pool area. This pattern aligns with expectations, as spatters display irregular and dynamic behaviors in thermal videos, whereas the melt pool movement is more predictable. In contrast, S-DDIM without physics simulation information predicts high variance in the meltpool area, suggesting that the model is unsure about the location of the meltpool. The comparison highlights the advantage of Algorithm~\ref{alg:sample} in predictive variance reduction. The predictive uncertainty derived from KOH does not provide meaningful insights. KOH+VAE
approach quantifies uncertainties better than standard KOH.
However, the uncertainty on the edge of the meltpool is
high, indicating that KOH+VAE is uncertain about the precise
geometry of the meltpool. The contrast further illustrates \name's ability to accurately characterize the distribution of the evolution of physical systems.}

\section{Conclusion and future research directions}
This paper introduces \name, a diffusion-based surrogate model for multi-fidelity physics simulations calibration. 
We envision that the flexibility of \name would engender broader applications in manufacturing and beyond. 

In the case studies presented in this paper, the parameters within the physics models are assumed known a priori or are calibrated (by nonlinear least square fitting) prior to deploying the \name model. Consequently, an intriguing avenue for future research would be to integrate parameter calibration directly within the probabilistic diffusion model framework. 

\bibliographystyle{IEEEtran}
\bibliography{reference.bib}

\clearpage
\newpage
\onecolumn

\appendices
\setcounter{page}{1}

\begin{center}
    \Large{Supplement for the paper:} \\
    \smallskip
    \Large{Diffusion-Based Surrogate Modeling and Multi-Fidelity Calibration} \\
    \smallskip
\end{center}

\bigskip

\section{Derivations in Algorithm~\ref{alg:gdef}}
In this section, we will discuss the intuitions for  Algorithm~\ref{alg:gdef} in the main paper. The goal is to derive an estimate for $\nabla \log p(\bc_{2,s^{*}}|\bx_{t,s^{*}},\bc_{1,s^{*}})$. The derivation is inspired by~\cite{chung2023guide}.

We first use the Bayes law to rewrite $\log p(\bc_{2,s^{*}}|\bx_{t,s^{*}},\bc_{1,s^{*}})$ as,
\begin{align*}
&\log p(\bc_{2,s^{*}}|\bx_{t,s^{*}},\bc_{1,s^{*}})=\log p(\bx_{t,s^{*}}|\bc_{1,s^{*}}) + \log \int p(\bc_{2,s^{*}}|\bx_{0,s^{*}},\bx_{t,s^{*}},\bc_{1,s^{*}})p(\bx_{0,s^{*}}|\bx_{t,s^{*}},\bc_{1,s^{*}})d\bx_{0,s^{*}}+C,
\end{align*}
where $C$ is a constant independent of $\bx_{0,s^{*}}$. 

As the integral can be difficult to calculate analytically, we resort to approximations. There are a few feasible approximations, among which we will use the most conceptually simple and computationally tractable one. Since we have trained a denoising network $\veceps_{\theta}(\cdot)$, we can approximate  $p(\bx_{0,s^{*}}|\bx_{t,s^{*}},\bc_{1,s^{*}})$ by Tweedie’s formula~\cite{chung2023guide},
\begin{equation*}
p(\bx_{0,s^{*}}|\bx_{t,s^{*}},\bc_{1,s^{*}})\approx \delta\left(\bx_{0,s^{*}}-\hat{\bx}_{0,s^{*},\theta}(\bx_{t,s^{*}},\bc_{1,s^{*}},t)\right),
\end{equation*}
where $\delta(\cdot)$ is the Dirichlet-delta function and $\hat{\bx}_{0,s^{*},\theta}(\bx_{t,s^{*}},\bc_{1,s^{*}},t)$ is defined as,
\begin{equation}
\hat{\bx}_{0,s^{*},\theta}(\bx_{t,s^{*}},\bc_{1,s^{*}},t)=  \frac{\bx_{t,s^{*}}-\sqrt{1-\bar{\alpha_t}}\veceps_{\theta}(\bx_{t,s^{*}},\bc_{1,s^{*}},t)}{\sqrt{\bar{\alpha_t}}}.  
\end{equation}

With the approximation, we can calculate the gradient as,
\begin{align*}
&\nabla_{\bx_{t,s}} \log p(\bc_{2,s^{*}}|\bx_{t,s^{*}},\bc_{2,s^{*}}) \approx  \nabla_{\bx_{t,s^{*}}} \log  p(\bc_{2,s^{*}}|\hat{\bx}_{0,s^{*},\theta}(\bx_{t,s^{*}},\bc_{1,s^{*}},t),\bx_{t,s^{*}},\bc_{1,s^{*}}).
\end{align*}

We can use the Leibniz rule to further expand the gradient as, 
$$\nabla_{\bx_{t,s^{*}}} \log  p(\bc_{2,s^{*}}|\bx_{t,s^{*}},\bx_{t,s^{*}},\bc_{1,s^{*}}) \approx \nabla_{\hat{\bx}_{0,s^{*},\theta}} \log  p(\bc_{2,s^{*}}|\hat{\bx}_{0,s^{*},\theta}(\bx_{t,s^{*}},\bc_{1,s^{*}},t),\bx_{t,s^{*}},\bc_{1,s^{*}})\frac{\partial \hat{\bx}_{0,s^{*},\theta}}{\partial \bx_{t,s^{*}}}. 
$$

The exact calculation of $\frac{\partial \hat{\bx}_{0,s^{*},\theta}}{\partial \bx_{t,s^{*}}}$ requires taking the gradient of a high dimensional variable through a neural network, which can be memory-consuming. To save memory, we thus employ another approximation by ignoring the gradient contributed by the denoising network, $\frac{\partial \hat{\bx}_{0,s^{*},\theta}}{\partial \bx_{t,s^{*}}}\approx \frac{1}{\sqrt{\bara}}\mathbf{I}$. Combing these two approximations, we have,
\begin{align*}
\nabla_{\bx_{t,s^{*}}} \log  p(\bc_{2,s^{*}}|\bx_{t,s^{*}}(\bx_{t,s^{*}},\bc_{1,s^{*}},t),\bx_{t,s},\bc_{1,s^{*}}) \approx \frac{1}{\sqrt{\bara_t}}\nabla_{\hat{\bx}_{0,s,\theta}} \log  p(\bc_{2,s^{*}}|\hat{\bx}_{0,s^{*},\theta},\bx_{t,s^{*}},\bc_{1,s^{*}}).
\end{align*}

\section{Derivations of the solution to the heat equation}
The Green's function solves the following equation
\begin{equation}
\label{eqn:greenfunctioncondition}
\ppt G = \nabla\cdot \left(\kappa \nabla G\right) - \rho G +\delta(s - s^{\prime})\delta(\br-\brp),
\end{equation}
where $\delta(\cdot)$ is the Dirichelet delta function.

We can use the Fourier transform to solve the equation. Due to translational invariance, the Fourier series of $G$ is given by,
\begin{equation}
G(\br,s;\brp,s^{\prime}) = \int\expn{i \bk^{\top}(\br-\brp)}\mg(\bk,s,s^{\prime})d\bk.
\end{equation}

Therefore, equation \eqref{eqn:greenfunctioncondition} becomes,
\begin{equation}
\label{eqn:mgequation}
 \ppt \mg = -(\bk^{\top}\kappa \bk +\rho )\mg + \delta(s - s^{\prime})\expn{-\bk^{\top}(\br-\brp)},   
\end{equation}
where we have used the fact $\delta(\br-\brp)=\int \expn{\bk^{\top}(\br-\brp)}d\bk$.

When $s>\spr$, the solution to \eqref{eqn:mgequation} is simply,
\begin{align*}
&\mg(\bk,s,s^{\prime}) = \expn{-\rho(s - s^{\prime})}\times\expn{-i\bk^{\top} (\br-\brp)-\bk^{\top}\kappa \bk (s - s^{\prime})}.
\end{align*}

Therefore, the original Green's function is 
\begin{equation*}
\begin{aligned}
&G(\br,s;\brp,s^{\prime}) = \expn{-\rho(s - s^{\prime})}\left(\frac{1}{\sqrt{2\pi}}\right)^{-1}\left(\frac{\det (\kappa^{-1})}{2(s - s^{\prime})}\right)^{-1}\expn{-\frac{(\br-\brp)\kappa^{-1}(\br-\brp)}{4(s - s^{\prime})}}.
\end{aligned}
\end{equation*}
It consists of a exponential decaying component $\expn{-\rho(s - s^{\prime})}$, a Gaussian component $\expn{-\frac{(\br-\brp)\kappa^{-1}(\br-\brp)}{4(s - s^{\prime})}}$, and a constant $C_n=\left(\frac{1}{\sqrt{2\pi}}\right)^{-1}\left(\frac{\det (\kappa^{-1})}{2(s - s^{\prime})}\right)^{-1}$.

\section{Optical flow and wrapping}
In this section, we first introduce the details of spatter dynamics modeling, then present our implementation of the wrapping operator.

\subsection{Optical flow modeling}
Remember that we assume spatters are transported by a velocity field $\bv(x,y,s)=[v_x,v_y]^{\top}\in\mathbb{R}^2$. Thus, in the ideal world, the spatter particle temperature field $u_p$ should satisfy,
\begin{align*}
u_p(x,y,s)=u_p(x-v_x\Delta s,y-v_y\Delta s,s-\Delta s).
\end{align*}

In practice, the optical flow may not be perfectly accurate because of the modeling, measurement, optimization, and statistical errors. We thus use the following probability model to characterize the inaccuracy,
\begin{equation}
u_p(x,y,s)=u_p(x-v_x\Delta s,y-v_y\Delta s,s-\Delta s) +\gamma^{-\frac{1}{2}}\varepsilon,
\end{equation}
where $\varepsilon$ are i.i.d. standard Gaussians noise. The parameter $\gamma$ represents our prior belief about the model accuracy: a larger $\gamma$ implies a higher confidence in the flow model. Therefore, given the optical flow $\bv$, the conditional probability is,
\begin{align}
\label{apeqn:continuousprob}
&\log p(\bv|u_p(\cdot,\cdot,s),u_p(\cdot,\cdot,s-\Delta s))=C-\frac{\gamma}{2}\times\notag\\
&\iint\left(u_p(x,y,s)-u_p(x-v_x\Delta s,y-v_y\Delta s,s-\Delta s)\right)^2dx dy.
\end{align}

\begin{figure}[h]
    \centering
    \includegraphics[width=0.6\textwidth]{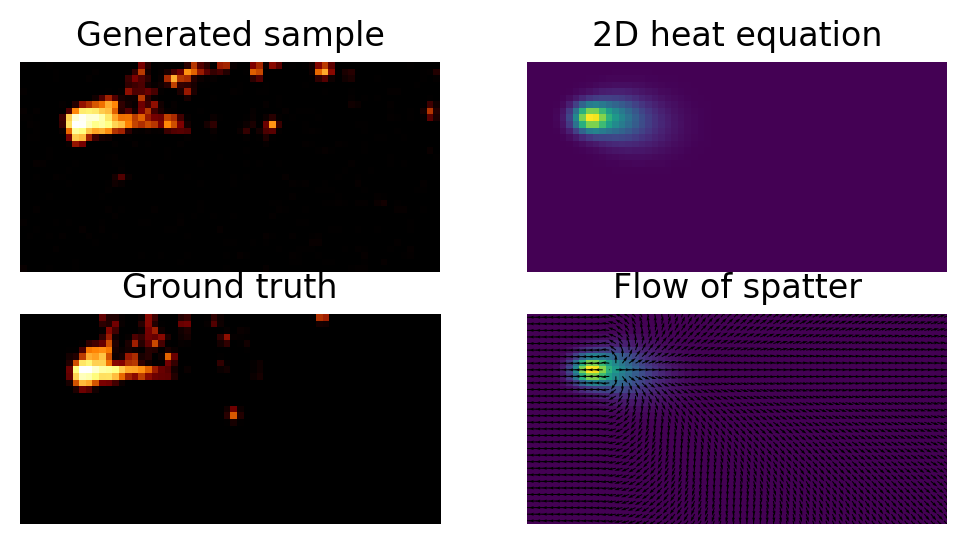}
    
    \caption{\revise{An illustration of a sample generated by \name, the corresponding $\bc_{s,1}$ (heat equation), $\bc_{s,2}$ (flow of spatter), and the ground truth. The directions of the flow of spatters are plotted as the small black arrows on the bottom-right figure.}}
    \label{fig:amepoch19}
\end{figure}

Since we predict the temperature only on discrete pixels rather than the continuous 2D plane, we should extend ~\eqref{apeqn:continuousprob} to the discrete counterpart defined on the $W\times H$ grids. First, we discretize the velocity field $\bv$ on the same $W \times H$ grid. An example of the discretized velocity field is shown as black arrows in the bottom-right graph of Figure\ref{fig:amepoch19}. Next, we use this discretized velocity field to link the temperature fields $u_p$ between two consecutive time steps. The resulting conditional probability is modeled by~\eqref{eqn:flowphysicsprobability}, where $\wrap(\cdot)$ represents the discrete version of $\bv$-transport.

\subsection{Wrapping operatior implementation}
We will introduce our implementation of $\wrap(\cdot)$ in the rest of this section. As discussed, we use $\bx_{0,s-1}$ to represent the vecterized temperature at time $s-1$ evaluated on the $W\times H$ grids. We also use $\bv$ to denote the velocity on the same grids. The wrapping operator uses $\bv$ and $\bx_{0,s-1}$ to predict $\bx_{0,s}$, which is the temperature at time $s$ defined on the $W\times H$ grids.

\begin{figure}[h]
    \centering 
    \caption{An illustration of the advection and the wrapping operator. Blue arrows denote velocity defined on grids. The orange dot is traced back in time.}
    \includegraphics[width=0.35\textwidth]{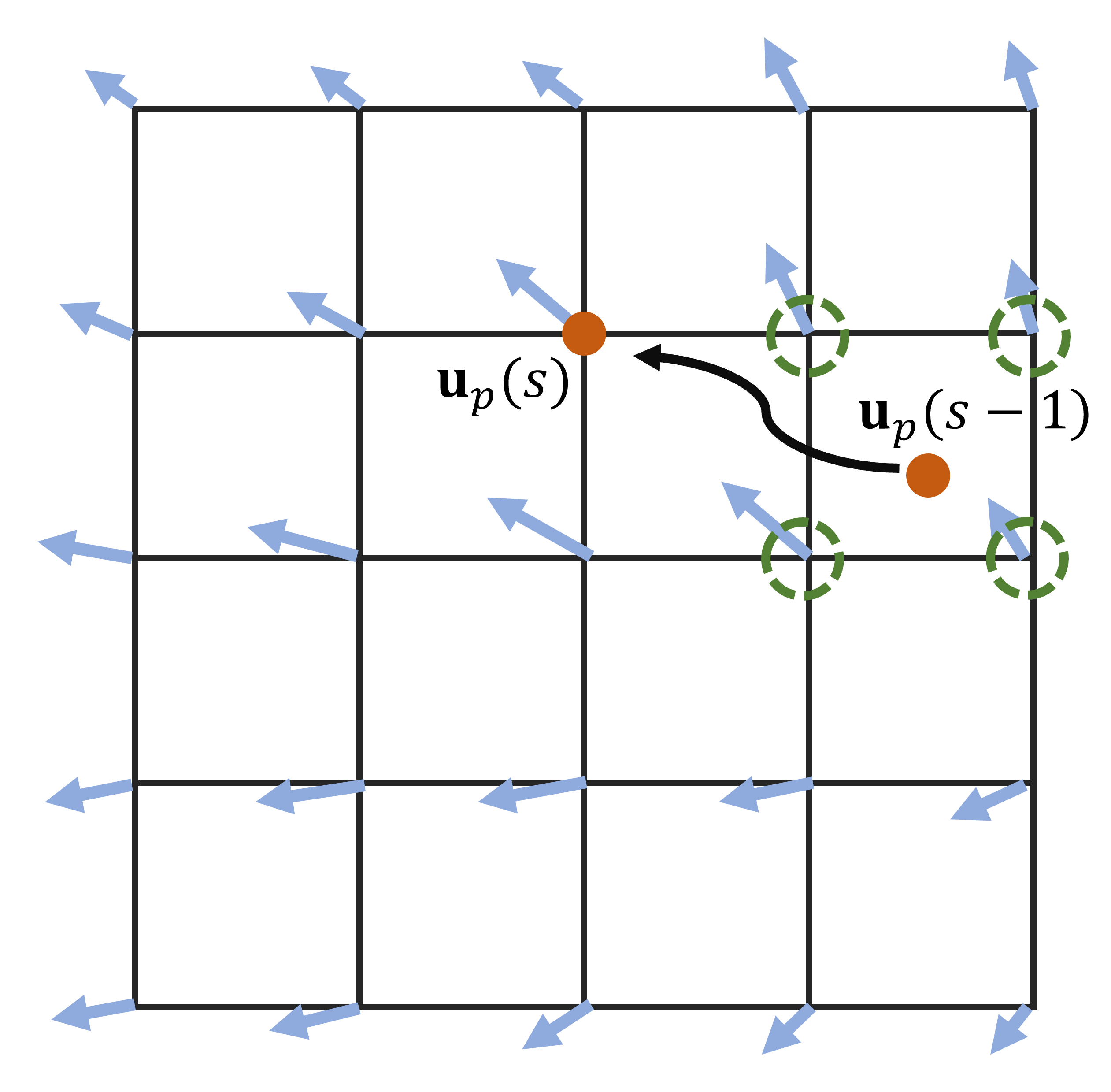}
    \label{fig:interpolate}
\end{figure}

We implement the wrapping operator via a semi-Lagrangian approach. An illustration of the numerical advection is shown in Fig.~\ref{fig:interpolate}.

More specifically, for any grid location $(x,y)$, the algorithm estimates its location at time $s-1$ as is $(x-v_x,y-v_y)$, then predict $u(x,y,t)$ as $u(x-v_x,y-v_y,s-1)$. However, as shown in Fig.~\ref{fig:interpolate}, the orange dot representing $(x-v_x,y-v_y)$ may not sit exactly on the grid point. Thus, $u(x-v_x,y-v_y,s-1)$ is not readily known. To resolve the issue, we use bilinear interpolation to interpolate the velocity $u(x-v_x,y-v_y,s-1)$ based on the grid velocity $\bx_{0,s-1}$ on the closest grids, which are shown as the green circles in Fig.~\ref{fig:interpolate}. 

We provide a pseudo-code for the wrapping operator in Algorithm~\ref{alg:wrap}.

\begin{algorithm}
\caption{The wrapping operator $\wrap$}
\label{alg:wrap}
\begin{algorithmic}[1]
\STATE Input $\bx_{0,s-1},v_x,v_y$ defined on $W\times H$ grids.

\FOR{
Row index $i=1,...,H$}
\FOR{Column index $j=1,\cdots,W$}
\STATE Calculate $(j-v_y(i,j),i-v_x(i,j))$.  
\STATE  Calculate $u(i,j,s) = \texttt{interpolate}\left(\bx_{0,s-1},j-v_y,i-v_x\right)$.
\ENDFOR
\ENDFOR
\STATE Calculate $\bx^{\texttt{pred}}_{0,s}$ by stacking values of $u(i,j,s)$.
\STATE Return $\bx^{\texttt{pred}}_{0,s}$.
\end{algorithmic}
\end{algorithm}

In Algorithm~\ref{alg:wrap}, \texttt{interpolate} is the standard bilinear interpolation from the closest grid points. The entire operation is differentiable.

\section{\revise{Additional visualizations and numerical simulations}}
\label{apsec:additionalexperiment}
\revise{In this section, we provide additional visualizations and numerical simulation results. First, we present two illustrations: one depicting the U-Net architecture and another showcasing the U-Net latent feature map. Next, we describe the auto-regressive procedures used to generate two sample videos. Finally, we present the results of an ablation study investigating the roles of $\bc_{s,1}$ and $\bc_{s,2}$ in the fluid simulation dataset.}

\subsection{U-net atructure}
\revise{The U-Net processes channel-wise concatenated input data using a series of 2D convolution, self-attention, and down-sampling modules to extract latent semantic features. During this process, the spatial resolution decreases from $128\times 128$ to $16\times 16$, while the number of channels increases to $1024$. Subsequently, the U-Net applies symmetric 2D convolution, self-attention, and up-sampling modules to reconstruct the score prediction from the extracted latent features.}

\revise{The time input 
$t$ is first transformed using a sine embedding module to create richer high-frequency information.  A fully connected
layer then maps these sine functions into time embedding vectors, which are integrated to the U-net main
framework by addition and multiplication.} 

\revise{We optimize the U-Net using the Adam optimizer in Algorithm~\ref{alg:trainepsilon}. Each denoising neural network (DeNN) is trained for $200$ epochs in the fluid simulation calibration task and for $100$ epochs in the laser-based additive manufacturing dataset, which is comparatively simpler.}

\begin{figure}[H]
    \centering
    \includegraphics[width=0.75\textwidth]{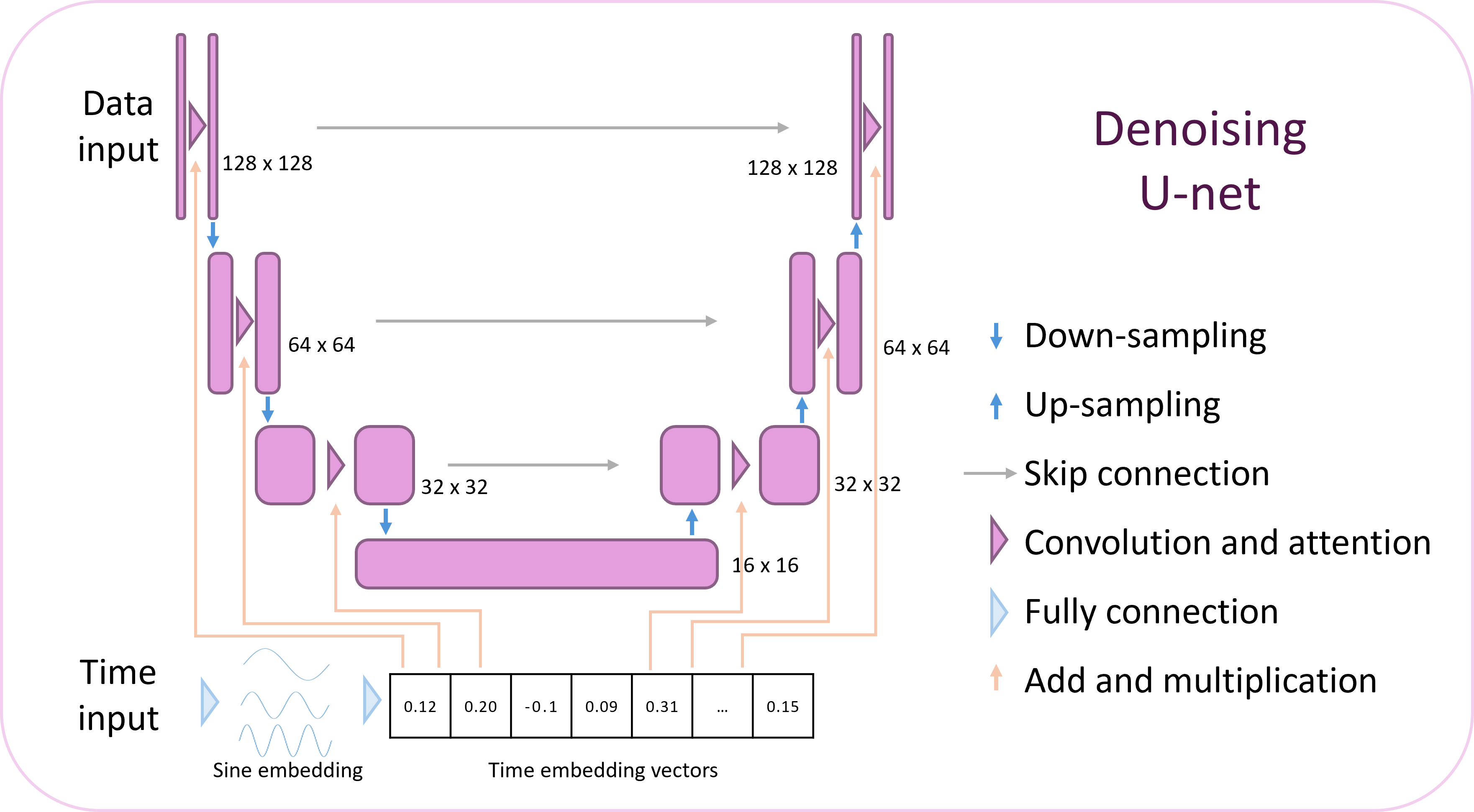}
    
    \caption{\revise{U-net architecture from~\cite{ddpm}. It consists of multiple 2D convolution, self-attention, down-sampling and up-sampling modules. Time embedding vectors are integrated to these operations through addition and multiplication.}}
    \label{fig:unetarchitecture}
\end{figure}

\subsection{U-net latent features}

\begin{figure}[H]
    \centering
    \includegraphics[width=0.95\textwidth]{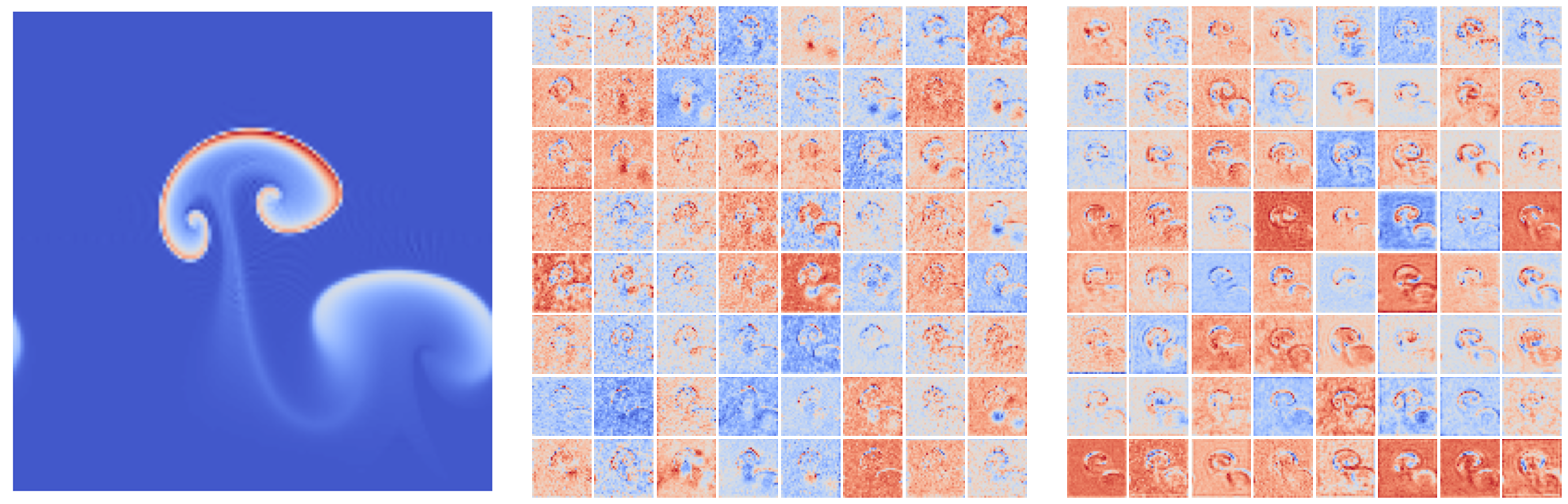}
    
    \caption{\revise{\textbf{Left}: the \name sample. \textbf{Middle}: latent features from Unet trained by standard DDIM. \textbf{Right}: latent features from Unet trained by \name with contextual input $\bc_{s,1}$. }}
    \label{fig:latentfeature}
\end{figure}

\revise{To further examine the latent features extracted by the U-Net, we visualize the feature maps of a 64-channel, $32\times 32$ representation in Figure~\ref{fig:latentfeature}.}

\revise{More specifically, we use \name to generate a sampling path, with the final generated sample shown on the left of Figure~\ref{fig:latentfeature}. We then introduce noise corresponding to $t=10$ and pass the resulting sample through the pre-trained U-Net to extract the latent features. The 64-channel features are arranged in an $8\times 8$ grid. For comparison, the middle panel of Figure~\ref{fig:latentfeature} presents the feature map of the sample processed by a U-Net trained with standard DDIM, while the right panel shows the feature map of the same sample obtained from a U-Net trained with contextual information $\bc_{s,1}$. }

\revise{From Figure~\ref{fig:latentfeature}, one can see that latent features from Unet trained by \name with contextual input are less noisy and focus more on the regions where buoyancy changes drastically. In contrast, the latent features in the middle are less visually informative. Such comparison highlights the benefits of using $\bc_{s,1}$ in helping the denoising neural networks to understand the semantic latent features, thus yielding better predictive performance.} 

\subsection{Two sample videos}

\revise{We generate two sample videos on the evolution of the fluid system and the laser-based manufacturing process. The links are presented below.}
\begin{itemize}
    \item Fluid system: {\color{blue} \url{https://drive.google.com/file/d/1XFU1TWuzYcQSe8w-xZzwlNVCvgBilm7R/view?usp=sharing}}.
    \item Additive manufacturing: {\color{blue} \url{https://drive.google.com/file/d/1yIKgi2Nnk0qDkoKppTT2WKXh0qSu4W6o/view?usp=sharing}}
\end{itemize}

\revise{To generate these videos, we first run both expensive and inexpensive physics simulations on a test setting not included in the training data. Next, we use fully-trained denoising neural networks (DeNNs) to calibrate their outputs by \name, refining the simulation results to improve fidelity.}

\subsection{Ablation study}
\revise{To further examine the contribution of $\bc_{s,1}$ and $\bc_{s,2}$, we conduct an ablation study. In addition to the results presented in Table~\ref{tab:fumes}, we train a DeNN without information from the inexpensive physics simulation. During the sampling stage, we employ a conditional diffusion process with the same energy-based guidance as described in Algorithm~\ref{alg:sample} with Choice 2. The evaluation metrics on the test set for this setting are reported in the third row of Table~\ref{tab:ablation}.}

\begin{table}[htbp]
  \centering
  \caption{The mean and standard deviation of the sample quality of different algorithms.}
    \begin{tabular}{ccccc}
    \toprule
          & MSE (0.001)$\downarrow$   & PSNR$\uparrow$  & SSIM $\uparrow$ & LPIPS$\downarrow$ \\
    \midrule
   
    S-DDIM & 3.69(0.2) & 25.5(0.2) & 99.955(0.001) & 0.401(0.002) \\
     \name with $\bc_{s,1}$  & \underline{1.14}(0.08) & \underline{33.2}(0.4) & \underline{99.992}(0.001) & 0.287(0.009) \\
     \name with $\bc_{s,2}$  & 1.12(0.06) & 32.1(0.3) & 99.992(0.001) & \textbf{0.216}(0.006) \\
    \name with $\bc_{s,1},\bc_{s,2}$  & \textbf{1.00}(0.05) & \textbf{33.6}(0.3) & \textbf{99.993}(0.001) & \underline{0.228}(0.006) \\    
    \bottomrule
    \end{tabular}%
  \label{tab:ablation}%
\end{table}%

\revise{In Table~\ref{tab:ablation}, the \name model with both inexpensive simulation $\bc_{s,1}$ and expensive simulation $\bc_{s,2}$ achieves the best performance in terms of MSE, PSNR, and SSIM. If we remove inexpensive physics $\bc_{s,1}$ and only include the guidance from expensive physics $\bc_{s,2}$, MSE increases, and PSNR and SSIM decrease. Since these metrics represent the pixel-wise resemblance between generated samples and the ground truth, the comparison suggests that $\bc_{s,1}$ is useful for the diffusion model to generate samples closer to the ground truth.}

\section{Proof for Theorem~\ref{thm:wassdis}}
In this section, we present the proof for Theorem~\ref{thm:wassdis} in the main paper.
In literature, the Wasserstein distance is defined as,
\begin{equation}
    \wdiv{p_1(\bx)}{p_2(\bx)} = \min_{\pi\in \Pi(p_1,p_2)}\left(\int\norm{\bx-\by}^2d\pi(\bx,\by)\right)^{\frac{1}{2}},
\end{equation}
where $\Pi(p_1.p_2)$ is the set of all joint distributions of $(\bx,\by)$ whose marginal distributions are $p_1(\bx)$ and $p_2(\by)$.

We will begin by introducing a useful inequality that provides an upper bound of the Wasserstein distance between the distributions given by two PDEs. Similar versions of this inequality have already been derived in literature~\cite{wassdiv}. 
\begin{lemma}
\label{lm:wassdis}
Consider two p.d.f. $p_{1,t}(\bx)$ and $p_{2,t}(\bx)$ in $\mathbb{R}^d$ that satisfy the following PDEs respectively,
\begin{align}
    \frac{\partial}{\partial t} p_{1,t}(\bx) +\nabla \cdot\left(p_{1,t}(\bx)\mu_1(\bx_t,t)\right)=0 \tag{PDE1}\label{apeqn:pde1},\\
    \frac{\partial}{\partial t} p_{2,t}(\bx) +\nabla \cdot\left(p_{2,t}(\bx)\mu_2(\bx,t)\right)=0 \tag{PDE2}\label{apeqn:pde2}.
\end{align}
where $\mu_1,\mu_2\in\mathbb{R}^d$ are drift terms that satisfy the regularity conditions in~\cite{wassdiv}. If additionally, there exists a finite function $L_1(t)$, such that for each $t\in[0,T]$, $\lvert\left(\bx-\by\right)^{\top}\left(\mu_2(\bx,t)-\mu_2(\by,t)\right)\rvert\le L_1(t)\norm{\bx-\by}^2,\, \forall \bx,\by\in \mathbb{R}^d$, then, the Wasserstain distance satisfies 
\begin{align*}
  \wdiv{p_{1,0}}{p_{2,0}} \le \wdiv{p_{1,T}}{p_{2,T}}\expn{\int_0^TL_1(r)dr} +\int_{0}^T\expn{\int_0^rL_1(r)dt}\mathbb{E}_{\bx\sim p_{1,t}(\bx)}\left[\norm{\mu_1(\bx,t)-\mu_2(\bx_t)}^2\right]^{\frac{1}{2}}dt.
\end{align*}
\end{lemma}

\revise{The proof of Lemma~\ref{lm:wassdis} follows~\cite{wassdiv}. A similar version is also presented in~\cite{stochasticinterpolant}.  We thus omit the complete proof for brevity.}

\revise{Intuitively, Lemma~\ref{lm:wassdis} bounds the Wasserstein distance between two p.d.f. $p_{1,0}$ and $p_{2,0}$ by the summation of the Wasserstein distance between $p_{1,T}$ and $p_{2,T}$ and the integrated distance of the drift terms in~\eqref{apeqn:pde1} and~\eqref{apeqn:pde1}. To prove the Theorem~\ref{thm:wassdis} in the main paper, we only need to estimate the differences between the two drift terms $\mu_1-\mu_2$. In the following, we will show that the difference is upper bounded by $\mathcal{L}_1$ and $\mathcal{L}_2$. Our proof slightly extends~\cite{wassdiv} as in~\eqref{eqn:expensivediffusionsample}, the reverse diffusion process involves a gradient guidance term~\eqref{eqn:term3}. }

The rest of this appendix presents the technical details in establishing the Wasserstein distance upper bounds. Throughout the discussion, we assume the regularity conditions in~\cite{wassdiv} are satisfied. Additionally, we assume that there are constants $L_{\epsilon},L_g>0$, such that for all possible tuple $(\bx_{0,1:\sct},\bc_{1,s^{*}},\bc_{2,s^{*}}, t)$, the following holds,
\begin{equation}
\left\{
\begin{aligned}
   \left\lvert\left(\bx-\by\right)^{\top}\left(\veceps_{\theta}(\bx,\bx_{0,1:\sct},\bc_{1,s^{*}})-\veceps_{\theta}(\by,\bx_{0,1:\sct},\bc_{1,s^{*}})\right)\right\rvert &\le L_{\epsilon}\norm{\bx-\by}^2\\
   \left\lvert\left(\bx-\by\right)^{\top}\left(g(\bx,\bc_{1,s^{*}},\bc_{2,s^{*}})-g(\by,\bc_{1,s^{*}},\bc_{2,s^{*}})\right)\right\rvert &\le L_g\norm{\bx-\by}^2
\end{aligned}
\right. \quad \forall \bx,\by\in \mathbb{R}^d
\end{equation}
\begin{proof}
We first consider the sampling algorithm with choice $1$. In the continuous regime, the forward diffusion process is characterized by
\begin{equation*}
d\bx_{t,s^{*}} = -\frac{\beta_t}{2}\bx_{t,s^{*}} dt+\sqrt{\beta_t}d\bw_t.
\end{equation*}

The forward Kolmogorov equation~\cite{cmumath} shows that the p.d.f. $p(\bx_{t,s^{*}}|\bx_{0,1:\sct},\bc_{1,s^{*}})$ follows a PDE,
\begin{equation*}
    \frac{\partial}{\partial t} p(\bx_{t,s^{*}}|\bx_{0,1:\sct},\bc_{1,s^{*}}) +\nabla \cdot\left(p(\bx_{t,s^{*}}|\bx_{0,1:\sct},\bc_{1,s^{*}})\left(-\frac{\beta_t}{2}\bx-\frac{\beta_t}{2}\nabla \log p(\bx_{t,s^{*}}|\bx_{0,1:\sct},\bc_{1,s^{*}})\right)\right)=0
\end{equation*}

As discussed, the sampling algorithm with choice 1 reduces to an ODE in the continuous regime,
\begin{equation}
\label{eqn:approxreverseprocess}
d\bx_{t,s^{*}} = \left(\frac{\beta_t}{2}\bx_{t,s^{*}}-\frac{\beta_t}{2\sqrt{1-\bara_t}}\veceps_{\theta}(\bx_{t,s^{*}},\bx_{0,1:\sct},\bc_{1,s^{*}})\right) 
\end{equation}
and initalizes $\bx_{T,s}$ from the standard normal distribution $\mathcal{N}(0,\mathbf{I})$. By forward Kolmogorov equation, the p.d.f. of $\bx_{s,t}$, $\qcun$, satisfy,
\begin{equation*}
    \frac{\partial}{\partial t} \qcun(\bx_{t,s^{*}}) +\nabla \cdot\left(\qcun(\bx_{t,s^{*}})\left(-\frac{\beta_t}{2}\bx_{t,s^{*}}+\frac{\beta_t}{2}\frac{\veceps_{\theta}(\bx_{t,s^{*}},\bx_{0,1:\sct},\bc_{1,s^{*}})}{\sqrt{1-\bara_t}}\right)\right)=0
\end{equation*}

It is straightforward to set $\mu_1(\bx_{t,s^{*}})=-\frac{\beta_t}{2}\bx_{t,s^{*}}+\frac{\beta_t}{2}\nabla \log p(\bx_{t,s^{*}}|\bx_{0,1:\sct},\bc_{1,s^{*}})$, $\mu_2(\bx_{t,s^{*}})=-\frac{\beta_t}{2}\bx_{t,s^{*}}-\frac{\beta_t}{2\sqrt{1-\bara_t}}\veceps_{\theta}(\bx_{t,s^{*}},\bx_{0,1:\sct},\bc_{1,s^{*}})$, and apply Lemma~\ref{lm:wassdis}. We first provide an upper bound of the Lipshitz constant $L_1$ from the assumptions. For any $ \bx,\by\in \mathbb{R}^d$, we have,
\begin{align*}
&\lvert\left(\bx-\by\right)^{\top}\left(\mu_2(\bx,t)-\mu_2(\by,t)\right)\rvert\\
&=\left\lvert\left(\bx-\by\right)^{\top}\left(\frac{\beta_t}{2}\left(\bx-\by\right)+\frac{\beta_t}{2\sqrt{1-\bara_t}}\left(\veceps_{\theta}(\bx)-\veceps_{\theta}(\by)\right)\right)\right\rvert\\
&\le \frac{\beta_t}{2}\norm{\bx-\by}^2 + \frac{\beta_t}{2\sqrt{1-\bara_t}}\lvert\left(\bx-\by\right)^{\top}\left(\veceps_{\theta}(\bx)-\veceps_{\theta}(\by)\right)\rvert\\
&\le \frac{1+L_{\epsilon}}{2}\norm{\bx-\by}^2,
\end{align*}
where we used the inequality $\frac{\beta_t}{2\sqrt{1-\bara_t}}\le \frac{\beta_t}{2\sqrt{\beta_t}}$ and $\beta_t<1$.

Therefore, Lemma~\ref{lm:wassdis} implies,
\begin{align*}
&\wdiv{p(\bx_{0,s^{*}}|\bx_{0,1:\sct,\bc_{1,s^{*}}})}{\qcun(\bx_{0,s^{*}})}\\
&\le \wdiv{p(\bx_{T,s^{*}}|\bx_{0,1:\sct,\bc_{1,s^{*}}})}{\mathcal{N}(0,\mathbf{I})}\expn{T\frac{1+L_{
\epsilon
}}{2}} \\
&+ \int_{0}^T\expn{t\frac{1+L_{
\epsilon
}}{2}}\frac{\beta_t}{2} \mathbb{E}_{p_t(\bx_{t,s^{*}})}\left[\norm{\nabla p(\bx_{t,s^{*}}|\bx_{0,1:\sct},\bc_{1,s^{*}})+\frac{\veceps_{\theta}(\bx_{t,s^{*}},\bx_{0,1:\sct},\bc_{1,s^{*}})}{\sqrt{1-\bara}}}^2\right]^{\frac{1}{2}}dt \\
&\le\wdiv{p(\bx_{T,s^{*}}|\bx_{0,1:\sct,\bc_{1,s^{*}}})}{\qcun(\bx_{T,s^{*}})}\expn{T\frac{1+L_{
\epsilon
}}{2}}+ \left(\int_{0}^T\expn{t(1+L_{
\epsilon
})}\frac{\beta_t}{2}dt\right)^{\frac{1}{2}} \times  \sqrt{\mathcal{L}_1}\\
\end{align*}
where we used Cauchy-Schwartz inequality in the last inequality and $\mathcal{L}_1$ is defined as,
\begin{equation*}
 \mathcal{L}_1 = \frac{1}{2}\int_0^T\beta_t\mathbb{E}_{p_t(\bx_{t,s^{*}})}\left[\norm{\nabla p(\bx_{t,s^{*}}|\bx_{0,1:\sct},\bc_{1,s^{*}})+\frac{\veceps_{\theta}(\bx_{t,s^{*}},\bx_{0,1:\sct},\bc_{1,s^{*}})}{\sqrt{1-\bara}}}^2\right]dt   .
\end{equation*}

Then, we proceed to analyze the sampling algorithm with choice 2. The ideal reverse process conditioned on $\bc_{1,s^{*}}$ and $\bc_{2,s^{*}}$ is given by,
\begin{align}
\label{eqn:cs2reverse}
d\bx_{t,s^{*}} &= \left(\frac{\beta_t}{2}\bx_{t,s^{*}}+\frac{\beta_t}{2}\nabla \log p(\bx_{t,s^{*}}|\bx_{0,1:\sct},\bc_{1,s^{*}},\bc_{2,s^{*}})\right) dt\notag\\
&=\left(\frac{\beta_t}{2}\bx_{t,s^{*}}+\frac{\beta_t}{2}\nabla \log p(\bx_{t,s^{*}}|\bx_{0,1:\sct},\bc_{1,s^{*}})+\frac{\beta_t}{2}\nabla \log p(\bc_{2,s^{*}}|\bx_{t,s^{*}},\bc_{1,s^{*}})\right) dt.
\end{align}
The corresponding Kolmogorov equation is given by,
\begin{align*}
    &\frac{\partial}{\partial t} p(\bx_{t,s^{*}}|\bx_{0,1:\sct},\bc_{1,s^{*}},\bc_{2,s^{*}}) \\
    &=\nabla \cdot\left(p(\bx_{t,s^{*}}|\bx_{0,1:\sct},\bc_{1,s^{*}},\bc_{2,s^{*}})\left(\frac{\beta_t}{2}\bx+\frac{\beta_t}{2}\nabla \log p(\bx_{t,s^{*}}|\bx_{0,1:\sct},\bc_{1,s^{*}})+\frac{\beta_t}{2}\nabla \log p(\bc_{2,s^{*}}|\bx_{t,s^{*}},\bc_{1,s^{*}})\right)\right).
\end{align*}

Similarly, the continuous form of the approximate reverse process employed by choice 2 of the sampling algorithm is
\begin{equation}
\label{eqn:cs2approxreverseprocess}
d\bx_{t,s^{*}} = \left(\frac{\beta_t}{2}\bx_{t,s^{*}}-\frac{\beta_t}{2\sqrt{1-\bara_t}}\veceps_{\theta}(\bx_{t,s^{*}},\bx_{0,1:\sct},\bc_{1,s^{*}})+\frac{\beta_t}{2}\bg(\bx_{0,1:\sct},\bc_{1,s^{*}},\bc_{2,s^{*}})\right) dt.
\end{equation}

Its corresponding Kolmogorov equation is,
\begin{align*}
    &\frac{\partial}{\partial t} \qcdeux(\bx_{t,s^{*}}) \\
    &=\nabla \cdot\left(\qcdeux(\bx_{t,s^{*}})\left(\frac{\beta_t}{2}\bx+\frac{\beta_t}{2\sqrt{1-\bara}}\nabla \veceps_{\theta}(\bx_{t,s^{*}},\bx_{0,1:\sct},\bc_{1,s^{*}})+\frac{\beta_t}{2}\nabla \bg(\bx_{0,1:\sct},\bc_{1,s^{*}},\bc_{2,s^{*}})\right)\right).
\end{align*}

Similarly, if we set $\mu_1(\bx_{t,s^{*}})=\frac{\beta_t}{2}\bx_{t,s^{*}}+\beta_t\nabla p(\bx_{t,s^{*}}|\bx_{0,1:\sct},\bc_{1,s^{*}})+\beta_t\nabla p(\bc_{2,s^{*}}|\bx_{t,s^{*}},\bc_{1,s^{*}})$ and $\mu_2(\bx_{t,s^{*}})=\frac{\beta_t}{2}\bx_{t,s^{*}}-\frac{\beta_t}{\sqrt{1-\bara_t}}\veceps_{\theta}+\beta_t\bg$, we can calculate $L_1(r)\le\frac{1+L_{\epsilon}+L_g}{2}$. As a result,  Lemma~\ref{lm:wassdis} indicates
\begin{align*}
&\wdiv{p(\bx_{0,s^{*}}|\bx_{0,1:\sct},\bc_{1,s^{*}},\bc_{2,s^{*}})}{\qcdeux(\bx_{0,s^{*}})}\\
&\le \wdiv{p(\bx_{T,s^{*}}|\bx_{0,1:\sct},\bc_{1,s^{*}},\bc_{2,s^{*}})}{\mathcal{N}(0,\mathbf{I})}\expn{T\frac{1+L_{
\epsilon
}+L_g}{2}}+ \left(\int_{0}^T\expn{t(1+L_{
\epsilon
}+L_g)}\frac{\beta_t}{2}dt\right)^{\frac{1}{2}}\\
&\times  \sqrt{\frac{1}{2}\int_0^T\beta_t\mathbb{E}_{p_t(\bx_{t,s^{*}})}\left[\norm{\nabla p(\bx_{t,s^{*}}|\bx_{0,1:\sct},\bc_{1,s^{*}})+\frac{\veceps_{\theta}(\bx_{t,s^{*}},\bx_{0,1:\sct},\bc_{1,s^{*}})}{\sqrt{1-\bara}}+\nabla \log p(\bc_{2,s^{*}}|\bx_{t,s^{*}},\bc_{1,s^{*}})-\nabla\bg}^2\right]dt}.
\end{align*}
From the inequality $\norm{a+b}^2\le 2\norm{a}^2+2\norm{b}^2$, we have,
\begin{align*}
 &\norm{\nabla p(\bx_{t,s^{*}}|\bx_{0,1:\sct},\bc_{1,s^{*}})+\frac{\veceps_{\theta}(\bx_{t,s^{*}},\bx_{0,1:\sct},\bc_{1,s^{*}})}{\sqrt{1-\bara}}+\nabla \log p(\bc_{2,s^{*}}|\bx_{t,s^{*}},\bc_{1,s^{*}})-\nabla\bg}^2  \\
 &\le 2\norm{\nabla p(\bx_{t,s^{*}}|\bx_{0,1:\sct},\bc_{1,s^{*}})+\frac{1}{\sqrt{1-\bara}}\veceps_{\theta}(\bx_{t,s^{*}},\bx_{0,1:\sct},\bc_{1,s^{*}})}^2+2\norm{\nabla p(\bc_{2,s^{*}}|\bx_{t,s^{*}},\bc_{1,s^{*}})-\bg}^2.
\end{align*}

Combining them, we know
\begin{align*}
&\wdiv{p(\bx_{0,s^{*}}|\bx_{0,1:\sct},\bc_{1,s^{*}},\bc_{2,s^{*}})}{\qcdeux(\bx_{0,s^{*}})}\\
&\le \wdiv{p(\bx_{T,s^{*}}|\bx_{0,1:\sct},\bc_{1,s^{*}},\bc_{2,s^{*}})}{\mathcal{N}(0,\mathbf{I})}\expn{T\frac{1+L_{
\epsilon
}+L_g}{2}}+ \left(\int_{0}^T\expn{t(1+L_{
\epsilon
}+L_g)}\frac{\beta_t}{2}dt\right)^{\frac{1}{2}}\times  \sqrt{2\mathcal{L}_1+2\mathcal{L}_2},
\end{align*}
where $\mathcal{L}_2$ is defined as,
\begin{equation*}
 \mathcal{L}_2 = \frac{1}{2}\int_0^T\beta_t\mathbb{E}_{p_t(\bx_{t,s^{*}})}\left[\norm{\nabla p(\bx_{t,s^{*}}|\bx_{0,1:\sct},\bc_{1,s^{*}}) -\bg(\bx_{0,1:\sct},\bc_{1,s^{*}},\bc_{2,s^{*}})}^2\right]dt   .
\end{equation*}
This completes our proof.
\end{proof}
\end{document}